\documentclass[journal]{IEEEtranTCOM}
\def\doublecolumn{1}

\normalsize

\usepackage[cmex10]{amsmath}  %cmex10 ensures that only type 1 fonts will be used
\usepackage{amssymb}
\usepackage{graphicx,color}   %color for pdf_t generated by xfig

\usepackage{ifxetex}
\ifxetex
\usepackage{xunicode}% unicode character macros
\else
\usepackage[utf8]{inputenc}
\usepackage[T1]{fontenc}
\fi

\usepackage{cite}  %combined consecutive references
\usepackage{subfigure}
\newtheorem{theorem}{Theorem}
\newtheorem{remark}{Remark}
\newtheorem{lemma}{Lemma}
\newtheorem{definition}{Definition}

\newtheorem{proposition}{Proposition}

\newtheorem{corollary}{Corollary}

\newcommand{\mbf}[1]{\ensuremath{\boldsymbol{#1}}}
\ifx\doublecolumn\undefined
\else
\fi

\begin{document}

%---------- Title ----------
\title{The Three-User  Finite-Field Multi-Way Relay Channel with Correlated Sources}
\author{Lawrence Ong, Gottfried Lechner, Sarah J.\ Johnson, and Christopher M.\ Kellett
\thanks{Part of the material in this paper was presented at the IEEE International Symposium on Information Theory, Saint Petersburg, July 31--August 5, 2011.}
\thanks{This research was supported under Australian Research Council's (ARC) Discovery Projects funding schemes (DP1093114 and DP120102123). Lawrence Ong is the recipient of an ARC Discovery Early Career Researcher Award  (DE120100246). Sarah Johnson and Christopher Kellett are recipients of ARC Future Fellowships (FT110100195 and FT110100746 respectively).}
}

\maketitle

\begin{abstract}
This paper studies the three-user finite-field multi-way relay channel, where the users exchange messages via a relay. The messages are arbitrarily correlated, and the finite-field channel is linear and is subject to additive noise of arbitrary distribution. The problem is to determine the minimum achievable source-channel rate, defined as channel uses per source symbol needed for reliable communication.
We combine Slepian-Wolf source coding and functional-decode-forward channel coding to obtain the solution for two classes of source and channel combinations. Furthermore, for correlated sources that have their common information equal their mutual information, we propose a new coding scheme to achieve the minimum source-channel rate.
\end{abstract}


\section{Introduction}
We study the three-user multi-way relay channel (MWRC) with correlated sources, where each user transmits its data to the other two users via a single relay, and where the users' messages can be correlated.  %For two classes of source/channel combinations, we obtain a complete characterization for reliable communication, i.e., the set of all achievable rates, defined as channel uses per source symbol. 
%Correlated sources are commonly found in multiple geographically distributed measurements of the same type, e.g., temperature. One application of the MWRC with correlated sources is communication via a satellite~\cite{wynerwolf02}.
% of weather stations via a satellite, where multiple stations obtain measurements of their respective local weather condition, and each station is to obtain the weather conditions at all other stations by communicating with a satellite.
The MWRC is a canonical extension of the extensively studied two-way relay channel (TWRC), where two users exchange data via a relay~\cite{knopp06,rankovwittneben06,rankovwittneben07,kattigollakota07,gunduztuncel08,schnurrstanczak08,cuihokliewer09,hogowdasun12,ikkiaissa12,liangjingaowong13}. Adding users to the TWRC can change the problem significantly~\cite{kimsmida10,ongkellettjohnson12it,gunduzyenergoldsmithpoor13}.
% In this paper, we study the three-user  MWRC in which each user is to decode the data from the two other users, and where there is no direct link among the users. Communication is carried out via a single relay.
The MWRC has been studied from the point of view of channel coding and source coding.

In channel coding problems, the sources are assumed to be independent, and the channel noisy. The problem is to find the capacity, defined as the region of all {\em achievable} channel rate triplets (bits per channel use at which the users can encode/send on average). % \emph{Achievable rate tuples} here refer to the tuple of the number of message bits (per channel use) the users can transmit such that all other users can reliably recover their intended messages. The challenge is to find the \emph{capacity region} which is the closure of all achievable rate tuples. 
%Though the capacity region of the general MWRC remains unknown, Gündüz et al.~\cite{gunduzyener09} obtained asymptotic capacity results for the high SNR and the low SNR regimes for the Gaussian MWRC, and Ong et al.~\cite{ongjohnsonkellett10cl,ongmjohnsonit11} derived the capacity region of the finite field MWRC, which is achieved by \emph{functional-decode-forward} channel coding.
For the Gaussian MWRC with independent sources, G\"und\"uz et al.~\cite{gunduzyenergoldsmithpoor13} obtained asymptotic capacity results for the high SNR and the low SNR regimes. For the finite-field MWRC with independent sources, Ong et al.~\cite{ongjohnsonkellett10cl,ongmjohnsonit11} constructed the {\em functional-decode-forward} coding scheme, and obtained the capacity region. For the general MWRC with independent sources, however, the problem remains open to date.

In source coding problems, the sources are assumed to be correlated, but the channel noiseless. The problem is to find the region of all {\em achievable} source rate triplets (bits per message symbol at which the users can encode/send on average). The source coding problem for the three-user MWRC was solved by Wyner et al.~\cite{wynerwolf02}, using \emph{cascaded Slepian-Wolf} source coding~\cite{slepianwolf73}. 

In this paper, we study both source and channel coding in the same network, i.e., transmitting correlated sources through noisy channels (cf.\ our recent work~\cite{timolechnerongjohnson12} on the MWRC with correlated sources and orthogonal uplinks). % We do not restrict ourselves to separate source-channel coding.
For most communication scenarios, the source correlation is fixed by the natural occurrence of the phenomena, and the channel is the part that engineers are ``unwilling or unable to change''~\cite{massey88}. Given the source and channel models, we are interested in finding the limit of how fast we can feed the sources through the channel. To this end, define {\em source-channel rate}~\cite{gunduzerkipgoldsmithpoor09} (also known as bandwidth ratio~\cite{jaingunduzkulkarni12}) as the average channel transmissions used per source tuple. Our aim is then to derive the minimum source-channel rate required such that each user can {\em reliably} and {\em losslessly} reconstruct the other two users' messages.

%The problem is to find the minimum rate, defined as the number of channel uses required per reliable exchange of message triplet (one message for each user).
In the multi-terminal network, it is well known that separating source and channel coding, i.e., designing them independently, is not always optimal (see, e.g., the multiple-access channel \cite{dueck81}). Designing good joint source-channel coding schemes is difficult, let alone finding an optimal one. G\"und\"uz et al.~\cite{gunduzerkipgoldsmithpoor09} considered a few networks with two senders and two receivers, and showed that source-channel separation is optimal for certain classes of source structure. In this paper, we approach the MWRC in a similar direction. We show that source-channel separation is optimal for three classes of source/channel combinations, by constructing coding schemes that achieve the minimum source-channel rate. %It is worth noting that two of the three classes can be completely specified by the source structure alone, i.e., the results hold for any finite-field channel. 

Recently, Mohajer et al.~\cite{mohajertiandiggavi10} solved the problem of linear deterministic relay networks with correlated sources. They constructed an optimal coding scheme, where each relay injectively maps its received channel output to its transmitted channel input. While this scheme is optimal for deterministic networks, such a scheme (e.g., the amplify-forward scheme in the additive white Gaussian noise channel) suffers from noise propagation in noisy channels and has been shown to be suboptimal for the MWRC with independent sources~\cite{gunduzyenergoldsmithpoor13}.

\section{Main Results}

\subsection{Source and Channel Models} \label{sec:model}

\ifx\doublecolumn\undefined
\begin{figure}[t]
\centering
\resizebox{10cm}{!}{\input{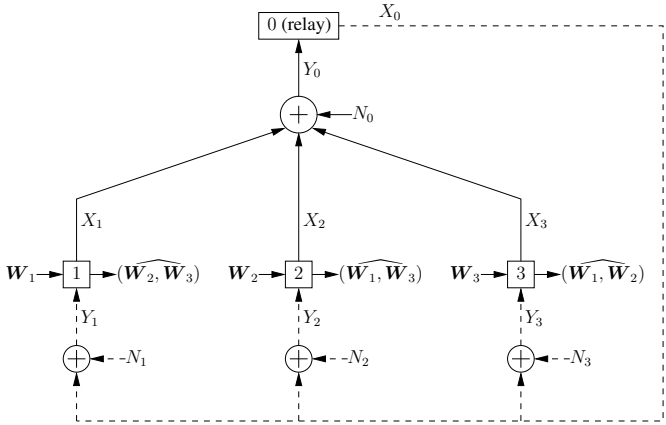}}
\caption{The three-user finite field MWRC with correlated sources: The uplink communications are represented by solid lines, and the downlink communications by dashed lines. The square blocks are nodes, and the circles represent finite field additions.}
\label{fig:mwrc}
\end{figure}
\else
\begin{figure}[t]
\centering
\includegraphics[width=\linewidth]{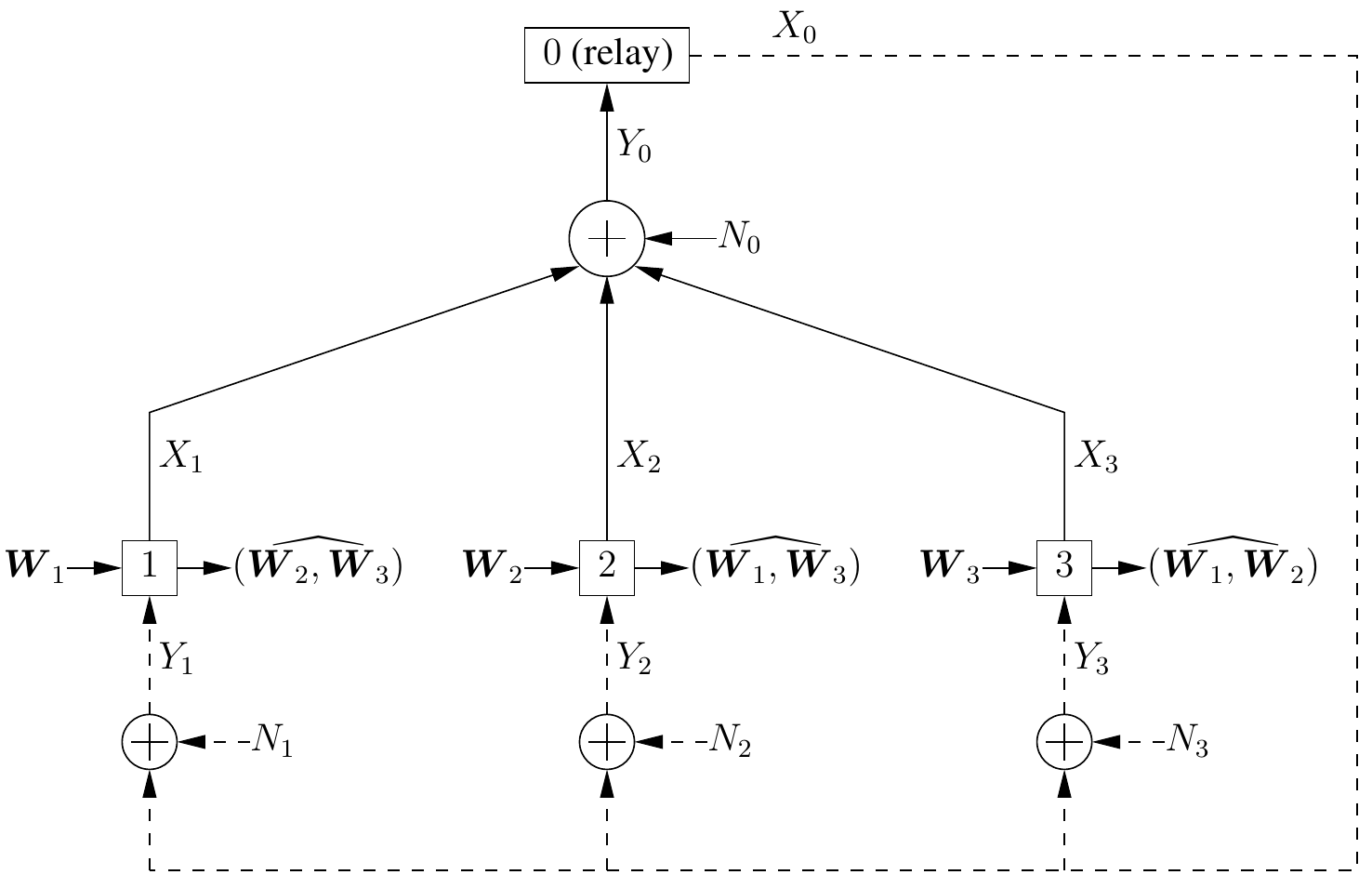}
\caption{The three-user finite field MWRC with correlated sources: The uplink communications are represented by solid lines, and the downlink communications by dashed lines. The square blocks are nodes, and the circles represent finite field additions.}
\label{fig:mwrc}
\end{figure}
\fi

We consider the MWRC depicted in Figure~\ref{fig:mwrc}, where three users (denoted by 1, 2, and 3) exchange messages through a noisy channel with the help of a relay (denoted by 0). For each node $d \in \{0,1,2,3\}$, we denote its source by $W_d$, its input to the channel by $X_d$, and its received channel output by $Y_d$. We let $W_0 = \varnothing$, as the relay has no source. 

We consider correlated and discrete-memoryless sources for the users, where $W_1$, $W_2$, and $W_3$ are generated according to some joint probability mass function
\begin{equation}
p(w_1,w_2,w_3). \label{eq:source}
\end{equation}
The channel consists of a finite-field {\em uplink} from the users to the relay, which takes the form
\begin{equation}
Y_0 = X_1 \oplus X_2 \oplus X_3 \oplus N_0, \label{eq:uplink}
\end{equation}
and a finite-field {\em downlink} from the relay to each user $i \in \{1,2,3\}$, which takes the form
\begin{equation}
Y_i = X_0 \oplus N_i,
\end{equation}
where $X_d, Y_d, N_d \in \mathcal{F}$, for all $d \in \{0,1,2,3\}$, for some finite field $\mathcal{F}$ of cardinality $|\mathcal{F}| = F$ with the associated addition $\oplus$. Here, $F$ can be any prime power. We assume that the noise $N_d$ is not uniformly distributed, i.e., its entropy $H(N_d) < \log F$; otherwise, it will randomize the channel, and no information can be sent through. 

Each user sends $m$ source symbols to the other two users (simultaneously) in $n$ channel uses. We refer to the $m$ source symbols of user $i$ as its message, denoted by $\mbf{W}_i \triangleq (W_i[1], W_i[2], \dotsc,$ $W_i[m])$, where each symbol triplet $(W_1[u], W_2[u], W_3[u])$ for $u \in \{1,2,\dotsc,m\}$ is generated independently according to \eqref{eq:source}. The channel is memoryless in the sense that the channel noise $N_d$ for all nodes and all channel uses are independent, and the distribution $p(n_d)$ is fixed for all channel uses. The source-channel rate, i.e., the number of channel uses per source triplet, is denoted by $\kappa \triangleq n/m$.

We assume that each user has all its $m$ source symbols prior to the $n$ channel uses\footnote{This assumption merely simplifies our analysis. Even if the source generation and the channel uses occur simultaneously ($m$ source triplets and $n$ channel uses per unit time), we can always transmit in blocks. We first wait for $m$ source triplets to be generated, and then use the channel $n$ times to transmit these source symbols (while waiting for the next $m$ triplets generation), and so on. Taking the number of blocks to be sufficiently large, the source-channel rate can be made as close to $n/m$ as desired.}, and consider the following block code of source-channel rate $n/m$:
\begin{enumerate}
\item The $t$-th transmitted channel symbol of each node $d$ depends on its message and its previously received channel symbols, i.e., $X_d[t] = f_{d,t} (\mbf{W}_d, Y_d[1], Y_d[2], \dotsc, Y_d[t-1])$, for all $d \in \{0,1,2,3\}$ and for all $t \in \{1,2,\dotsc, n\}$.% Here, $\mbf{W}_0 \triangleq \varnothing$, as the relay has no message to send.
\item Each user $i$ estimates the messages of the other users from its own message and all its received channel symbols, i.e., user $i$ decodes the messages from users $j$ and $k$ as $(\widehat{\mbf{W}_j,\mbf{W}_k}) = h_i (\mbf{Y}_i,\mbf{W}_i)$, for all distinct $i,j,k \in \{1,2,3\}$. We denote $\mbf{Y}_i \triangleq (Y_i[1],Y_i[2],$ $\dotsc,Y_i[n])$.\footnote{The length of a bold-faced vector, either $m$ for source symbols or $n$ for channel symbols is clear from context.} %We always define $j < k$ for convenience. 
\end{enumerate}

Note that utilizing \emph{feedback} is permitted in our system model. This is commonly referred to as the {\em unrestricted} MWRC (cf.\ the {\em restricted} MWRC \cite{rankovwittneben06,oechteringschnurr08,gunduztuncel08,gunduzyenergoldsmithpoor13}). We will see later that for the classes of source/channel combinations for which we find the minimum source-channel rate, feedback is not used. This means that feedback provides no improvement to source-channel rate for these cases. 

User $i$ makes a decoding error if $(\widehat{\mbf{W}_j,\mbf{W}_k}) \neq (\mbf{W}_j,\mbf{W}_k)$. We define $P_\text{e}$ as the probability that one or more users make a decoding error, and say that source-channel rate  $\kappa \triangleq n/m$ is \emph{achievable} if the following is true: for any $\zeta>0$, there exists at least one block code of source-channel rate $\kappa$ with $P_\text{e} < \zeta$. The aim of this paper is to find the infimum of achievable source-channel rates, denoted by $\kappa^*$. For the rest of the paper, we refer to $\kappa^*$ as the minimum source-channel rate. %In other words, we want to find the minimum number of channel uses per reliable exchange of source triplet.

\begin{remark}
Theoretical interest aside, the finite-field channel considered in this paper shares two important properties with the AWGN channel (commonly used to model wireless environments). Firstly, the channel is linear, i.e., the channel output is a function of the sum of all inputs. Secondly, the noise is additive. Sharing these two properties, optimal coding schemes derived for the finite-field channel shed light on how one would code in AWGN channels. For example, the optimal coding scheme derived for the finite-field MWRC with independent sources~\cite{ongjohnsonkellett10cl} is used to prove capacity results for the AWGN MWRC with independent sources~\cite{ongkellettjohnson12it}.
\end{remark}

\subsection{Main Results} \label{sec:main-results}

We will now state the main result of this paper. The technical terms (in italics) in the theorem will be defined in Section~\ref{sec:definition} following the theorem.
\begin{theorem} \label{theorem:main}
The minimum source-channel rate is given by
\begin{equation}
\kappa^* = \max_{\text{distinct } i,j,k} \left\{ \frac{H(W_j,W_k|W_i)}{\log F - \max \{ H(N_0), H(N_i)\} } \right\},\footnote{Note that $\log F - \max \{ H(N_0), H(N_i)\} > 0$ by definition, since the noise at each receiver is not uniformly distributed.} \label{eq:main-result}
\end{equation}
if the sources have any one of the following:
\begin{enumerate}
\item {\em almost-balanced conditional mutual information}, or
\item {\em skewed conditional entropies} (on any {\em symmetrical} finite-field channel), or
\item {\em their common information equals their mutual information}.
\end{enumerate}
\end{theorem}

For Cases 1 and 2, we derive the achievability (upper bound) of $\kappa^*$ using existing (i) Slepian-Wolf source coding and (ii) functional-decode-forward channel coding for independent sources. We abbreviate this pair of source and channel coding scheme by SW/FDF-IS. We derive a lower bound using cut-set arguments. While the achievability for these two cases is rather straightforward, what we find interesting is that using the scheme for independent messages is actually optimal for two classes of source/channel combinations. %More surprisingly, the first class (Case 1) poses restrictions on only the sources. This means for this class of sources, this scheme is optimal for all (finite-field) channel.
Furthermore, although the source-channel rates achievable using SW/FDF-IS cannot be expressed in a closed form, we are able to derive closed-form conditions for two classes of sources where the achievability of SW/FDF-IS matches the lower bound.

In SW/FDF-IS, the source coding---while compressing---destroys the correlation among the sources, and hence channel coding for independent sources is used. For Case 3, the sources have their common information equal their mutual information, meaning that each source is able to identify the parts of the messages it has in common with other source(s). For this case, we again use Slepian-Wolf source coding, but we conserve the parts that the sources have in common. We then design a {\em new} channel coding scheme that takes the common parts into account. Here, the challenge is to optimize the functions of different parts that the relay should decode. We show that the new coding scheme is able to achieve $\kappa^*$.

For all three cases, the coding schemes are derived based on the separate source-channel coding architecture. Also, for Cases 1 and 3, $\kappa^*$ is found when only the sources satisfy certain conditions, and this is true independent of the underlying finite-field channel, i.e., any $F$ and any noise distribution.

\subsection{Definitions} \label{sec:definition}

In this section, we define the technical terms in Theorem~\ref{theorem:main}.

\subsubsection{Symmetrical Channel}

%\subsubsection{\underline{Symmetrical channels}} \label{sec:symmetrical}

\begin{definition} \label{def:symmetrical}
A finite-field MWRC is {\em symmetrical} if
\begin{equation}
H(N_1) = H(N_2) = H(N_3).
\end{equation}
Otherwise, we say that the channel is \emph{asymmetrical}.
\end{definition}

%Here, $H(X)$ is the entropy of the random variable $X$.
We can think of $H(N_i)$ as the noise level on the downlink from the relay to user $i$. So, a symmetrical channel requires that the downlinks from the relay to all the users are equally noisy.
We do not impose any condition on the uplink noise level, $H(N_0)$. %, equals the noise level on the downlinks. % must equal those at the users. This means all downlinks from the relay to the users are equally noisy, but the uplink from the users to the relay can be noisier or less noisy. %An asymmetric channel is one which is not symmetric, i.e., different channels may be subject to different levels of noise.

\subsubsection{Almost-Balanced Conditional Mutual Information}

\begin{definition} \label{def:abcmi}
The sources are said to have \emph{almost-balanced conditional mutual information} (ABCMI) if
% \begin{multline}
% I(W_i;W_j|W_k) \leq I(W_j;W_k|W_i)+ I(W_i;W_k|W_j), \\ \forall i,j,k \in \{1,2,3\} \text{ and } i \neq j \neq k; \label{eq:balanced}
% \end{multline}
\begin{equation}
I(W_i;W_j|W_k) \leq I(W_j;W_k|W_i)+ I(W_i;W_k|W_j), \label{eq:balanced}
\end{equation}
for all distinct $i,j,k \in \{1,2,3\}$. Otherwise, the sources  are said to have \emph{unbalanced conditional mutual information}. 
\end{definition}

%Here, $I(X;Y|Z)$ is the conditional mutual information of the random variables $X$ and $Y$ given $Z$.

%So, if the sources have unbalanced conditional mutual information, there must exist some user $A \in \{1,2,3\}$, such that
Putting it another way, for unbalanced sources, we can always find a user $A \in \{1,2,3\}$, such that
% \begin{multline}
% I(W_B;W_C|W_A) = I(W_A;W_B|W_C)\\ + I(W_A;W_C|W_B) + \eta, \label{eq:unbalanced}
% \end{multline}
\begin{equation}
I(W_B;W_C|W_A) = I(W_A;W_B|W_C) + I(W_A;W_C|W_B) + \eta, \label{eq:unbalanced}
\end{equation}
for some $\eta > 0$ and distinct  $B,C \in \{1,2,3\}\setminus \{A\}$.

%Note that if \eqref{eq:balanced} fails, then there necessarily exists an appropriate positive $\eta$ for \eqref{eq:unbalanced}.

%The next definition divides the class of unbalanced conditional mutual information into two sub-classes.

\subsubsection{Skewed Conditional Entropies}

\ifx\doublecolumn\undefined
\begin{definition} \label{def:sce}
Sources with unbalanced conditional mutual information are said to have \emph{skewed conditional entropies} (SCE) if, in addition to \eqref{eq:unbalanced},
\begin{equation}
H(W_B,W_C|W_A)  \geq \max \Big\{ H(W_A,W_B|W_C) , H(W_A,W_C|W_B) \Big\} + \eta, \label{eq:case2a}
\end{equation}
for the same $\eta$ as in \eqref{eq:unbalanced}.
\end{definition}
\else
\begin{definition} \label{def:sce}
Sources with unbalanced conditional mutual information are said to have \emph{skewed conditional entropies} (SCE) if, in addition to \eqref{eq:unbalanced},
\begin{multline}
H(W_B,W_C|W_A)  \geq \max \Big\{ H(W_A,W_B|W_C) ,\\ H(W_A,W_C|W_B) \Big\} + \eta, \label{eq:case2a}
\end{multline}
for the same $\eta$ as in \eqref{eq:unbalanced}.
\end{definition}
\fi

%Here, $H(X,Y|Z)$ is the conditional entropy.

\subsubsection{Common Information Equals Mutual Information}

Lastly, we define {\em common information} in the same spirit as G{\'a}cs and K{\"o}rner~\cite{gacskorner72}. For two users, G{\'a}cs and K{\"o}rner defined common information as a value on which two users can {\em agree} (using the terminology of Witsenhausen~\cite{witsenhausen75}). The common information between two random variables can be as large as mutual information (in the Shannon sense), but no larger. %In general, common information is far smaller than common information. %  The maximum common information any two variables can have is the mutual information (in the Shannon sense); in general, common information is less than mutual information.

The concept of common information was extended to multiple users by Tyagi et al.~\cite{tyaginrayangupta11}, where they considered a value on which all users can agree. In this paper, we further extend common information to values on which different subsets of users can agree.
We now formally define a class of sources, where their common information equals their mutual information.

\begin{definition}\label{def:cc}
Three correlated random variables $(W_1, W_2, W_3)$ are said to have their common information equal their mutual information if there exists four random variables $V_{12}$, $V_{23}$, $V_{13}$, and $V_{123}$ such that
\begin{align}
V_{12} &= \phi_{12}(W_1) = \phi_{21}(W_2), \\
V_{23} &= \phi_{23}(W_2) = \phi_{32}(W_3), \\
V_{13} &= \phi_{13}(W_1) = \phi_{31}(W_3), \\
V_{123} &= \phi_{123}(W_1) = \phi_{213}(W_2) = \phi_{312}(W_3),
\end{align}
for some deterministic functions $\phi_\cdot(\cdot)$, and
\ifx\doublecolumn\defined
\begin{align}
H(V_{12}) &= I(W_1;W_2), \\
H(V_{23}) &= I(W_2;W_3), \\
H(V_{13}) &= I(W_1;W_3), \\
H(V_{123}) &= I(W_1;W_3) - I(W_1;W_3|W_2) \triangleq  I(W_1;W_2;W_3).
\end{align}
\else
\begin{align}
H(V_{12}) &= I(W_1;W_2), \\
H(V_{23}) &= I(W_2;W_3), \\
H(V_{13}) &= I(W_1;W_3), \\
H(V_{123}) &= I(W_1;W_3) - I(W_1;W_3|W_2) \nonumber \\ & \triangleq  I(W_1;W_2;W_3).
\end{align}
\fi
\end{definition}

We give graphical interpretations using information diagrams for sources that have ABCMI and SCE in Appendix~\ref{appendix:abcmi}, and examples of sources that have ABCMI and their common information equal their mutual information in Section~\ref{section:conclusion}.

% \begin{remark}
% For simplicity, we consider correlated sources with common cores in the sense of Han~\cite{han79}. Theorem~\ref{theorem:main} also holds for correlated sources in the sense of G{\'a}cs and K{\"o}rner~\cite{gacskorner72}.
% \end{remark}

\begin{figure}[t]
\centering
\includegraphics[width=7.9cm]{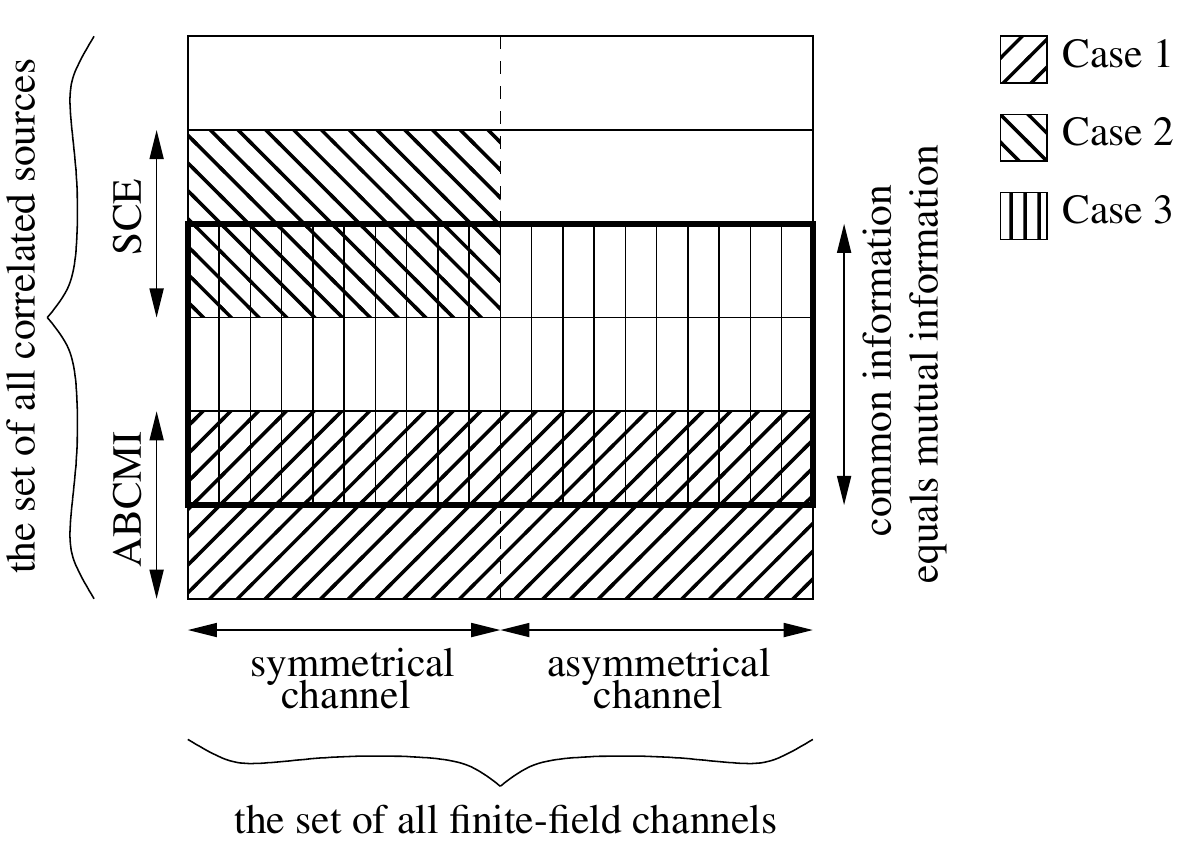}
\caption{Main results of this paper: shaded regions are the classes of source and channel combinations where the minimum source-channel rate is found}
\label{fig:results}
\end{figure}

%\begin{remark}
Definitions~\ref{def:abcmi} and \ref{def:sce} are mutually exclusive, but Definitions~\ref{def:cc} and \ref{def:abcmi} (or \ref{def:cc} and \ref{def:sce}) are not. This means correlated sources that have their common information equal their mutual information must also have either ABCMI, SCE, or unbalanced mutual information without SCE\@.
%Whether the sources have ABCMI or SCE is determined by the relative sizes of their conditional mutual information and conditional entropies. On the contrary, for the sources to have common cores, they must have a specific structure. For this class of sources, we do not impose any restriction on the entropies of $P$, $Q$, $R$, $S$, $T$, $U$, and $V$---the sources can have ABCMI or SCE.
This leads to the graphical summary of the results of Theorem~\ref{theorem:main} in Figure~\ref{fig:results}. 
%\end{remark}

\subsection{Organization}

The rest of this paper is organized as follows: We show a lower bound and an upper bound (achievability) to $\kappa^*$ in Section~\ref{section:lower}. In Section~\ref{section:upper-1-2}, we show that for Cases 1 and 2 in Theorem~\ref{theorem:main}, the lower bound is achievable. In Section~\ref{section:upper-3}, we propose a coding scheme that takes common information into account, and show the  source-channel rate achievable using this new scheme matches the lower bound. We conclude the paper with some discussion in Section~\ref{section:conclusion}.

\section{Lower and Upper  Bounds to $\kappa^*$} \label{section:lower}

Denote the RHS of \eqref{eq:main-result} as
\begin{equation}
\Phi \triangleq \max_{\text{distinct } i,j,k} \left\{ \frac{H(W_j,W_k|W_i)}{\log F - \max \{ H(N_0), H(N_i)\} } \right\}.
\end{equation}

We first show that $\Phi$ is a lower bound to $\kappa^*$.
Using cut-set arguments~\cite[pp.\ 587--591]{coverthomas06}, we can show that if source-channel rate $\kappa$ is achievable, then
\ifx\doublecolumn\undefined
\begin{subequations}
\begin{align}
H(W_j,W_k|W_i) &\leq \kappa \min \Big\{ I(X_j,X_k;Y_0,Y_i|X_0,X_i), I(X_j,X_k,X_0; Y_i|X_i) \Big\} \label{eq:upper-1} \\
&\leq \kappa \Big[ \log_2 F - \max \{ H(N_0), H(N_i)\} \Big], \label{eq:upper-2}
\end{align}
\end{subequations}
\else
\begin{subequations}
\begin{align}
H(W_j,W_k|W_i) &\leq \kappa \min \Big\{ I(X_j,X_k;Y_0,Y_i|X_0,X_i),\nonumber \\ &\quad\quad\quad\quad\;\; I(X_j,X_k,X_0; Y_i|X_i) \Big\} \label{eq:upper-1} \\
&\leq \kappa \Big[ \log_2 F - \max \{ H(N_0), H(N_i)\} \Big], \label{eq:upper-2}
\end{align}
\end{subequations}
\fi
for all distinct $i,j,k \in \{1,2,3\}$. Here \eqref{eq:upper-1} follows from  Mohajer et al.~\cite[eqs.\ (11)--(12)]{mohajertiandiggavi10} and \eqref{eq:upper-2} follows from Ong et al.~\cite[Section III]{ongmjohnsonit11}. Re-arranging the equation gives the following lower bound to all achievable source-channel rates $\kappa$---and hence also to $\kappa^*$:
\begin{lemma} \label{lemma:lower-bound}
For any three-user finite-field MWRC with correlated sources, the minimum source-channel rate is lower bounded as
\begin{equation}
\kappa^* \geq \Phi. \label{eq:lower-bound}
\end{equation}
\end{lemma}

We now present the result of SW/FDF-IS coding scheme that first uses Slepian-Wolf source coding for the noiseless MWRC with correlated sources~\cite{cover75}, followed by  functional-decode-forward for independent sources (FDF-IS) channel coding for the MWRC~\cite{ongmjohnsonit11}. This scheme achieves the following source-channel rates:
\begin{lemma} \label{lemma:achievability-separate}
For any three-user finite-field MWRC with correlated sources, SW/FDF-IS achieves all source-channel rates in $\mathcal{K}$, where 
\ifx\doublecolumn\undefined
\begin{equation}
\mathcal{K} = \left\{ \kappa \in \mathbb{R} \middle\vert
\begin{aligned}
& \kappa > \max_{\text{distinct } i,j,k} \left\{ \frac{H(W_i|W_j,W_k)}{R_i}, \frac{H(W_i,W_j|W_k)}{R_i + R_j} \right\},\\
& \text{for some positive } R_1, R_2, R_3 \text{ such that} \\
& R_j+R_k < \log F - \max \{ H(N_0), H(N_i)\}, \\
& \text{for all distinct } i,j,k \in \{1,2,3\}.
\end{aligned}
\right\}, \label{eq:achievability-separate}
\end{equation}
where $\mathbb{R}$ is the set of real numbers. So, the minimum source-channel rate is upper bounded as
\begin{equation}
\kappa^* \leq \inf \mathcal{K}.
\end{equation}
\else
\begin{equation}
\mathcal{K} = \left\{ \kappa \in \mathbb{R} \middle\vert
\begin{aligned}
& \kappa > \max_{\text{distinct } i,j,k} \left\{ \frac{H(W_i|W_j,W_k)}{R_i}, \right. \\ &\quad\quad\quad\quad\quad\quad\;\; \left.\frac{H(W_i,W_j|W_k)}{R_i + R_j} \right\},\\
& \text{for some positive } R_1, R_2, R_3 \text{ such that} \\
& R_j+R_k < \log F - \max \{ H(N_0), H(N_i)\}, \\
& \text{for all distinct } i,j,k \in \{1,2,3\}.
\end{aligned}
\right\}, \label{eq:achievability-separate}
\end{equation}
where $\mathbb{R}$ is the set of real numbers. So, the minimum source-channel rate is upper bounded as
\begin{equation}
\kappa^* \leq \inf \mathcal{K}.
\end{equation}
\fi
\end{lemma}

The proof is based on random coding arguments and can be found in Appendix~\ref{appendix:slepian-wolf-fdf}.

\begin{remark}
The variables $R_i$ are actually the channel code rates, i.e., the number of message bits transmitted by user $i$ per channel use.
\end{remark}

From Lemmas~\ref{lemma:lower-bound} and \ref{lemma:achievability-separate}, we have the following result:

\begin{corollary} \label{corollary:separation}
For a three-user finite-field MWRC, if $\inf \mathcal{K} = \Phi$, then $\kappa^* = \Phi = \inf \mathcal{K}$, meaning that the minimum source-channel rate is known and is achievable using SW/FDF-IS.
\end{corollary}

\begin{remark} \label{remark:separation}
The collection of source/channel combinations that satisfy Corollary~\ref{corollary:separation} forms a class where the minimum source-channel rate is found, in addition to Theorem~\ref{theorem:main}.
The challenge, however, is to characterize---in closed form---classes of source/channel combinations for which $\inf \mathcal{K} = \Phi$. For this, we need to guarantee the existence of three positive numbers $R_1, R_2,$ and $R_3$ satisfying the inequalities in \eqref{eq:achievability-separate} for every $\kappa > \Phi$.
\end{remark}

Next, we will show that for Cases 1 and 2 in Theorem~\ref{theorem:main}, SW/FDF-IS achieves all source-channel rates $\kappa > \Phi$.

\section{Proof of Cases 1 and 2 in Theorem~\ref{theorem:main}} \label{section:upper-1-2}

\subsection{Proof of Case 1 in Theorem~\ref{theorem:main}}

In this subsection, we will show that if the sources have ABCMI, then $\inf \mathcal{K} = \Phi$. Since any $\kappa \in \mathcal{K}$ relies on the existence of channel code rates $(R_1,R_2,R_3)$, we first show the following proposition:

\begin{proposition} \label{proposition:existence}
Consider sources with ABCMI\@. Given any source-channel rate $\kappa > 0$, and any positive number $\delta > 0$, we can always find positive $R_1, R_2,$ and $R_3$ such that 
\begin{align}
\kappa R_1 &\geq H(W_1|W_2,W_3) + \frac{\delta}{4}, \label{eq:lemma-case-1-a}\\
\kappa R_2 &\geq H(W_2|W_1,W_3) + \frac{\delta}{4}, \label{eq:lemma-case-1-b} \\
\kappa  R_3 &\geq H(W_3|W_1,W_2) + \frac{\delta}{4}, \label{eq:lemma-case-1-c} \\
\kappa (R_1+R_2) &= H(W_1,W_2|W_3) + \frac{\delta}{2}, \label{eq:lemma-case-1-d} \\
\kappa  (R_1+R_3) &= H(W_1,W_3|W_2) + \frac{\delta}{2}, \label{eq:lemma-case-1-e} \\
\kappa  (R_2+R_3) &= H(W_2,W_3|W_1) + \frac{\delta}{2}. \label{eq:lemma-case-1-f}
\end{align}
\end{proposition}

\begin{IEEEproof}[Proof of Proposition~\ref{proposition:existence}]
It can be shown that choosing
\ifx\doublecolumn\undefined
\begin{equation}
\kappa  R_i = H(W_i|W_j,W_k) + \frac{1}{2} \left[ I(W_i;W_k|W_j) + I(W_i;W_j|W_k) - I(W_j;W_k|W_i) \right] + \frac{\delta}{4}, \label{eq:case-1-r1-r3}
\end{equation}
\else
\begin{align}
\kappa  R_i &= H(W_i|W_j,W_k) + \frac{1}{2} \Big[ I(W_i;W_k|W_j) \nonumber \\ &\quad + I(W_i;W_j|W_k) - I(W_j;W_k|W_i) \Big] + \frac{\delta}{4}, \label{eq:case-1-r1-r3}
\end{align}
\fi
for all distinct $i,j,k \in \{1,2,3\}$ satisfies \eqref{eq:lemma-case-1-a}--\eqref{eq:lemma-case-1-f}. The expression in the square brackets is non-negative due to the ABCMI condition \eqref{eq:balanced}.
\end{IEEEproof}

With this result, we now prove Case 1 of Theorem~\ref{theorem:main}. We need to show that any source-channel rate $\kappa > \Phi$ is achievable, i.e., the source-channel rate
\begin{equation}
\kappa = \max_{\text{distinct } i,j,k} \left\{ \frac{H(W_j,W_k|W_i) + \delta}{\log F - \max \{ H(N_0), H(N_i)\} } \right\}, \label{eq:rate+delta}
\end{equation}
for any $\delta > 0$, lies in $\mathcal{K}$. Here, $\delta$ is independent of $F, W_i,$ and $N_i$.

For a source-channel rate in \eqref{eq:rate+delta}, we choose $\{R_i\}$ as in \eqref{eq:case-1-r1-r3}. Substituting \eqref{eq:lemma-case-1-d}--\eqref{eq:lemma-case-1-f} into \eqref{eq:rate+delta}, the second inequality in \eqref{eq:achievability-separate} is satisfied. Also, \eqref{eq:lemma-case-1-a}--\eqref{eq:lemma-case-1-f} imply the first inequality in \eqref{eq:achievability-separate}. Hence,  $\kappa \in \mathcal{K}$. This proves Case 1 in Theorem~\ref{theorem:main}. $\hfill \blacksquare$

% \begin{remark}
% If the sources are independent, we have $H(W_i,W_j|W_k) = H(W_i) + H(W_j)$ for all $i,j,k \in \{1,2,3\}$ where $i \neq j \neq k$. In this case, \eqref{eq:balanced} is always satisfied, i.e., all independent sources have ABCMI\@. Define $\tau_i = H(W_i)/\kappa = mH(W_i)/n$ as the number of message bits transmitted by user $i$ per channel use. Reliable communication is possible (i) only if $\tau_j + \tau_k \leq \log_2 |\mathcal{F}| - \max \{ H(N_0), H(N_i) \}$, for all $i,j,k \in \{1,2,3\}$ where $i \neq j \neq k$; and (ii) if the conditions hold with strict inequalities. With this we recover the capacity region of the three-user MWRC with independent sources~\cite{ongmjohnsonit11}.
% \end{remark};

\subsection{Proof of Case 2 in Theorem~\ref{theorem:main}}

We need to show that if the sources have SCE and the channel is symmetrical, then the source-channel rate in \eqref{eq:rate+delta} is achievable for any $\delta > 0$.
Recall that sources that have SCE must have unbalanced conditional mutual information, for which we can always re-index the users as $A$, $B$, and $C$ satisfying \eqref{eq:unbalanced} for some fixed $\eta>0$.

For achievability in Lemma~\ref{lemma:achievability-separate}, we first show the existence of $(R_A,R_B,R_C)$ satisfying the following conditions:
\begin{proposition} \label{proposition:existence-2}
Consider sources with unbalanced mutual information. Given any source-channel rate $\kappa > 0$, and any positive number $\delta > 0$, we can always find positive $R_A, R_B,$ and $R_C$ such that 
\begin{align}
\kappa R_A &= H(W_A|W_B,W_C) + \frac{\delta}{4}, \label{eq:lemma-case-2-a}\\
\kappa R_B &> H(W_B|W_A,W_C) + \frac{\eta}{2} + \frac{\delta}{4}, \label{eq:lemma-case-2-b} \\
\kappa R_C &> H(W_C|W_A,W_B) + \frac{\eta}{2} + \frac{\delta}{4}, \label{eq:lemma-case-2-c} \\
\kappa (R_A+R_B) &= H(W_A,W_B|W_C) + \frac{\eta}{2} + \frac{\delta}{2}, \label{eq:lemma-case-2-d} \\
\kappa (R_A+R_C) &= H(W_A,W_C|W_B) + \frac{\eta}{2} + \frac{\delta}{2}, \label{eq:lemma-case-2-e} \\
\kappa (R_B+R_C) &= H(W_B,W_C|W_A) + \frac{\delta}{2}, \label{eq:lemma-case-2-f}
\end{align}
for $\eta>0$ defined in \eqref{eq:unbalanced}.
\end{proposition}

\begin{IEEEproof}[Proof of Proposition~\ref{proposition:existence-2}]
Constraint \eqref{eq:unbalanced} implies the following:
\ifx\doublecolumn\undefined
\begin{align}
&I(W_B;W_C|W_A) + I(W_A;W_B|W_C) - I(W_A;W_C|W_B) = 2 I(W_A;W_B|W_C) + \eta > \eta, \label{eq:eta-1} \\
&I(W_B;W_C|W_A) + I(W_A;W_C|W_B) - I(W_A;W_B|W_C) = 2 I(W_A;W_C|W_B) + \eta > \eta. \label{eq:eta-2}
\end{align}
\else
\begin{align}
&I(W_B;W_C|W_A) + I(W_A;W_B|W_C) - I(W_A;W_C|W_B) \nonumber \\ &\quad = 2 I(W_A;W_B|W_C) + \eta > \eta, \label{eq:eta-1} \\
&I(W_B;W_C|W_A) + I(W_A;W_C|W_B) - I(W_A;W_B|W_C) \nonumber \\ &\quad = 2 I(W_A;W_C|W_B) + \eta > \eta. \label{eq:eta-2}
\end{align}
\fi

First, we can always choose a positive number $R_A$ as in \eqref{eq:lemma-case-2-a}.
In addition, we choose
\ifx\doublecolumn\undefined
\begin{align}
\kappa R_B &= H(W_B|W_A,W_C) + \frac{1}{2} \Big[ I(W_B;W_C|W_A) + I(W_A;W_B|W_C) - I(W_A;W_C|W_B)\Big] + \frac{\delta}{4}, \label{eq:case-2-r2}\\
\kappa R_C &= H(W_C|W_A,W_B) + \frac{1}{2} \Big[ I(W_B;W_C|W_A) + I(W_A;W_C|W_B) - I(W_A;W_B|W_C)\Big] + \frac{\delta}{4}. \label{eq:case-2-r3}
\end{align}
\else
\begin{align}
\kappa R_B &= H(W_B|W_A,W_C) + \frac{1}{2} \Big[ I(W_B;W_C|W_A)\nonumber \\ &\quad  + I(W_A;W_B|W_C) - I(W_A;W_C|W_B)\Big] + \frac{\delta}{4}, \label{eq:case-2-r2}\\
\kappa R_C &= H(W_C|W_A,W_B) + \frac{1}{2} \Big[ I(W_B;W_C|W_A)\nonumber \\ &\quad  + I(W_A;W_C|W_B) - I(W_A;W_B|W_C)\Big] + \frac{\delta}{4}. \label{eq:case-2-r3}
\end{align}
\fi

Substituting \eqref{eq:eta-1} into \eqref{eq:case-2-r2}, we get \eqref{eq:lemma-case-2-b}; substituting \eqref{eq:eta-2} into \eqref{eq:case-2-r3}, we get \eqref{eq:lemma-case-2-c}.
Summing different pairs from \eqref{eq:lemma-case-2-a}, \eqref{eq:case-2-r2}, and \eqref{eq:case-2-r3}, we get \eqref{eq:lemma-case-2-d}--\eqref{eq:lemma-case-2-f}.
\end{IEEEproof}

Furthermore, for a symmetrical channel, we can define
\begin{equation}
H_\text{downlink} \triangleq H(N_A) = H(N_B) = H(N_C), \label{eq:common-downlink-noise}
\end{equation}
So, \eqref{eq:case2a} for SCE and \eqref{eq:common-downlink-noise} for symmetrical channels imply that the source-channel rate in \eqref{eq:rate+delta} equals
\begin{equation}
\kappa = \frac{H(W_B,W_C|W_A) + \delta}{\log F - \max\{ H(N_0), H_\text{downlink} \} }. \label{eq:rate+delta+sce}
\end{equation}
Hence, we only need to show that the source-channel rate \eqref{eq:rate+delta+sce} is achievable for any $\delta>0$.

We first choose $R_A, R_B,$ and $R_C$ as in \eqref{eq:lemma-case-2-a}, \eqref{eq:case-2-r2}, and \eqref{eq:case-2-r3}, respectively. From \eqref{eq:lemma-case-2-d}--\eqref{eq:lemma-case-2-f}, we get
\begin{subequations}
\begin{align}
\kappa (R_B + R_C) & = H(W_B,W_C|W_A) +  \frac{\delta}{2} \label{eq:a} \\
& = \kappa \Big[\log F - \max\{ H(N_0), H_\text{downlink} \} \Big] - \frac{\delta}{2} \label{eq:b} \\
&  < \kappa \Big[\log F - \max\{ H(N_0), H_\text{downlink} \} \Big], \label{eq:2a-1}\\
\kappa (R_A + R_B) & = H(W_A,W_B|W_C) + \frac{\eta}{2} + \frac{\delta}{2} \label{eq:c} \\
& \leq H(W_B,W_C|W_A)  - \frac{\eta}{2} + \frac{\delta}{2} \label{eq:d}\\
& <  \kappa \Big[\log F - \max\{ H(N_0), H_\text{downlink} \} \Big], \label{eq:2a-2}\\
\kappa (R_A + R_C) & = H(W_A,W_C|W_B) + \frac{\eta}{2} + \frac{\delta}{2} \label{eq:e}\\
&  \leq H(W_B,W_C|W_A)  - \frac{\eta}{2} + \frac{\delta}{2} \label{eq:f} \\
&<   \kappa \Big[\log - \max\{ H(N_0), H_\text{downlink} \} \Big], \label{eq:2a-3}
\end{align}
\end{subequations}
where \eqref{eq:d} and \eqref{eq:f} follow from \eqref{eq:case2a}; \eqref{eq:b}, \eqref{eq:2a-2}, and \eqref{eq:2a-3} follow from \eqref{eq:rate+delta+sce}.\footnote{Note that $\eta$, which is determined by the sources' correlation, is strictly greater than zero; see its definition in \eqref{eq:unbalanced}.} This means the second inequality in \eqref{eq:achievability-separate} is satisfied. From \eqref{eq:lemma-case-2-a}--\eqref{eq:lemma-case-2-f}, we know that the first inequality in \eqref{eq:achievability-separate} is also satisfied. Hence, the source-channel rate \eqref{eq:rate+delta+sce} is indeed achievable for any $\delta>0$. $\hfill \blacksquare$

\subsection{A Numerical Example Showing that SW/FDF-IS is Not Always Optimal}

In this section, we give an example showing that SW/FDF-IS can be suboptimal. Consider the following sources: $W_1 = (U_1, U_{12}, U_{13})$, $W_2 = (U_2, U_{12}, U_{23})$, and $W_3 = (U_3, U_{13}, U_{23})$, where $U_1$ is uniformly distributed in $\{1,2,\dotsc, 2^6\}$, $U_{23}$ is uniformly distributed in $\{1,2,\dotsc, 2^3\}$, and $U_2, U_3, U_{12}, U_{13}$ are each uniformly distributed in $\{1,2\}$. In addition, all $U_i$ and $U_{ij}$ are mutually independent. Here, each $U_{ij}$ represents common information between $W_i$ and $W_j$.

For the channel, let the finite field be $\mathcal{F} = \{0,1,\dotsc, 2^{10}-1\}$ and $\oplus$ be modulo-$2^{10}$ addition, i.e., $F = 2^{10}$. Furthermore, let $\Pr\{N_0 = a\} = 2^{-2}$ for $a \in \{0,1,2,3\}$, and $\Pr\{N_0=a\} = 0$ for $a \in \mathcal{F} \setminus \{0,1,2,3\}$; let $\Pr\{N_i = a\} = 2^{-1}$ for $a \in \{0,1\}$, and  $\Pr\{N_i = a\} = 0$ for $a \in \mathcal{F} \setminus \{0,1\}$, for all $i \in \{1,2,3\}$.

For this source and channel combinations, we have $H(W_1|W_2,W_3) = 6$, $H(W_2|W_1,W_3) = H(W_3|W_1,W_2) = 1$, $H(W_1,W_2|W_3)=$ $H(W_1,W_3|W_2) =8$, $H(W_2,W_3|W_1) =5$, $\log F - H(N_0) = 8$, $\log F - H(N_i) = 9$, for all $i \in \{1,2,3\}$. One can verify that these sources have unbalanced conditional mutual information and do not have SCE.

In this example, $\Phi= 1$. Suppose that $\kappa = 1.05$ is achievable using SW/FDF-IS. From Lemma~\ref{lemma:achievability-separate}, there must exists three positive real numbers $R_1$, $R_2$, and $R_3$ such that
\ifx\doublecolumn\undefined
\begin{align}
1.05 &> \max \left\{ 6/R_1, 1/R_2, 1/R_3, 8/(R_1+R_2), 5/(R_2+R_3), 8/(R_1+R_3) \right\}, \label{eq:contradiction-a}\\
R_1 + R_2 &< \min \{ 8,9 \} = 8, \label{eq:contradiction-b}\\
R_2 + R_3 &< 8, \\
R_1 + R_3 &< 8. \label{eq:contradiction-c}
\end{align}
\else
\begin{align}
1.05 &> \max \big\{ 6/R_1, 1/R_2, 1/R_3, 8/(R_1+R_2),\nonumber \\ &\quad\quad\quad\quad  5/(R_2+R_3), 8/(R_1+R_3) \big\}, \label{eq:contradiction-a}\\
R_1 + R_2 &< \min \{ 8,9 \} = 8, \label{eq:contradiction-b}\\
R_2 + R_3 &< 8, \\
R_1 + R_3 &< 8. \label{eq:contradiction-c}
\end{align}
\fi
From \eqref{eq:contradiction-a}, we must have that $R_1 > 6/1.05$ and $R_2 + R_3 > 5/1.05$. These imply $\max \{R_1 + R_2, R_1 + R_3\} = R_1 + \max \{ R_2,R_3\} \geq R_1 + \frac{R_2+R_3}{2} >  8.5/1.05 = 8.095$. Hence, \eqref{eq:contradiction-b} and \eqref{eq:contradiction-c} cannot be simultaneously true. This means the source-channel rate 1.05 is not achievable using SW/FDF-IS\@.

The sources described here have their common information equal their mutual information. We will next propose an alternative scheme that is optimal for this class of sources. The following proposed scheme achieves all source-channel rates $\kappa >1$ for this source/channel combination, meaning that the minimum source-channel rate for this example is $\kappa^* =1$. So SW/FDF-IS is strictly suboptimal for this source/channel combination.

\section{Proof of Case 3 in Theorem~\ref{theorem:main}} \label{section:upper-3}

While the achievability for Cases 1 and 2 uses existing source and channel coding schemes, for Case 3 (i.e., sources that have their common information equal their mutual information), we will use an existing source coding scheme and design a new channel coding scheme to achieve all source-channel rates $\kappa > \Phi$.

\begin{remark}
While Case 3 in general requires a new achievability scheme, these sources may have ABCMI (as shown in Figure~\ref{fig:results}). For such cases, optimal codes can also be obtained using the coding scheme for Case 1.
\end{remark}

In this section, without loss of generality,\footnote{We can always re-index the users such that \eqref{eq:order} is true.} we let
\begin{equation}
H(W_1,W_2|W_3) \geq \max \{ H(W_2,W_3|W_1) , H(W_1,W_3|W_2) \}. \label{eq:order}
\end{equation}
This means we can re-write $\Phi$ as follows:
\begin{equation}
\Phi = \max \left\{ \frac{H(W_1,W_2|W_3)}{\log F - H(N_0)}, \max_{\text{distinct } i,j,k} \frac{H(W_j,W_k|W_i)}{\log F - H(N_i)} \right\}. \label{eq:kappa-simplified}
\end{equation}

%We will show that if the sources have only common information, rates arbitrarily close to $\kappa^*$ are achievable.

As mentioned earlier, we will use a separate-source-channel-coding architecture, where we first perform source coding and then channel coding. We will again use random coding arguments. More specifically, we will use random linear block codes for channel coding.

\subsection{Source Coding}

%We perform source coding on the $m$-vectors $\boldsymbol{W}_i$ for all $i \in \{1,2,3\}$. More specifically, we 
We encode each $\boldsymbol{V}_\mathcal{S} \in (V_\mathcal{S}[1], V_\mathcal{S}[2], \dotsc, V_\mathcal{S}[m])$ to $\boldsymbol{c}_\mathcal{S}$, which is a length-$\ell_\mathcal{S}$ finite-field (of size $F$) vector, for all $\mathcal{S} \in \{(12), (23), (13), (123)\}$ (see Definition~\ref{def:cc} for the definition of $V_\mathcal{S}$). We also encode each $\boldsymbol{W}_i$ to $\boldsymbol{c}_i$, which is a length-$\ell_i$ finite-field vector.  So, each message $\boldsymbol{W}_i$ is encoded into four subcodes, e.g., $\boldsymbol{W}_1$ is encoded into $(\boldsymbol{c}_1,\boldsymbol{c}_{12}, \boldsymbol{c}_{13}, \boldsymbol{c}_{123})$. Some subcodes---the common parts---are shared among multiple sources.

Using the results of distributed source coding~\cite{slepianwolf73,cover75}, if $m$ is sufficiently large and if
\begin{align}
\frac{\ell_i \log F}{m} &> H(W_i|W_j,W_k), && \textnormal{ for all } i, \label{eq:subcode-length-1}\\
%\frac{\ell_{ij} \log F}{m} &> I(W_i;W_j|W_k),\\
\frac{\ell_{ij} \log F}{m} &> I(W_i;W_j|W_k), && \textnormal{ for all } i < j, \label{eq:subcode-length-3} \\
\frac{\ell_{123} \log F}{m} &> I(W_1;W_2;W_3), \label{eq:subcode-length-4}
\end{align}
then we can decode $(\boldsymbol{c}_1,\boldsymbol{c}_{12}, \boldsymbol{c}_{13}, \boldsymbol{c}_{123})$ to $\boldsymbol{W}_1$, $(\boldsymbol{c}_2,\boldsymbol{c}_{12}, \boldsymbol{c}_{23}, \boldsymbol{c}_{123})$ to $\boldsymbol{W}_2$, and $(\boldsymbol{c}_3,\boldsymbol{c}_{13}, \boldsymbol{c}_{23}, \boldsymbol{c}_{123})$ to $\boldsymbol{W}_3$ with an arbitrarily small error probability. We show the proof in Appendix~\ref{app:GK-source-coding}.

After source coding, user 1 has $(\boldsymbol{c}_1, \boldsymbol{c}_{12}, \boldsymbol{c}_{13}, \boldsymbol{c}_{123})$. In order for it to decode $(\boldsymbol{W}_2,\boldsymbol{W}_3)$, it must receive $(\boldsymbol{c}_2, \boldsymbol{c}_3, \boldsymbol{c}_{23})$ from the other users through the channel. Similarly, users 2 and 3 must each obtain subcodes that they do not already have through the channel.

In contrast to the source coding used for Cases 1 and 2, here, we have generated source codes where the users share some subcodes. So, instead of using existing FDF-IS channel codes (designed for independent sources), we will design channel codes that take the common subcodes into account.

\begin{table*}[t]
\centering
\caption{Uplink transmission using linear block codes when $\ell_{12} \leq \ell_3$, $\ell_\text{T} \triangleq \ell_1 + \ell_{12} + \ell_2$}
\label{table:1}
\begin{tabular}{| c||c | c | c | c | c | c | c |}
\hline
Message length &  $\ell_{23}$ & $\ell_1 + \ell_{12} - \ell_3 - \ell_{23}$  & $\ell_3 - \ell_{12}$  & $\ell_{12}$ & $\ell_3 - \ell_{12}$ & $\ell_2 + \ell_{12} - \ell_3 - \ell_{13}$ & $\ell_{13}$  \\
\hline
Messages & $\mbf{c}_1^{(1)}$ & $\mbf{c}_1^{(3)}$ & $\mbf{c}_1^{(2)}$ & $\mbf{c}_{12}$ & $\mbf{c}_2^{(2)}$ & $\mbf{c}_2^{(3)}$ & $\mbf{c}_2^{(1)}$ \\
\hline
Messages &$\mbf{c}_{23}$ &  & $\mbf{c}_3^{(2)}$ & $\mbf{c}_3^{(1)}$ & $\mbf{c}_3^{(2)}$ &   & $\mbf{c}_{13}$ \\
\hline
Channel uses & $\frac{\ell_{23}}{\ell_\text{T}}n$ & $\frac{\ell_1 + \ell_{12} - \ell_3 - \ell_{23}}{\ell_\text{T}}n$ & $\frac{\ell_3 - \ell_{12}}{\ell_\text{T}}n$ & $\frac{\ell_{12}}{\ell_\text{T}}n$ & $\frac{\ell_3 - \ell_{12}}{\ell_\text{T}}n$ & $\frac{\ell_2 + \ell_{12} - \ell_3 - \ell_{13}}{\ell_\text{T}}n $ & $\frac{\ell_{13}}{\ell_\text{T}}n$\\
\hline
Total channel uses & \multicolumn{7}{c|}{$n$} \\
\hline
\end{tabular}
\end{table*}

\subsection{Channel Coding}

%Suppose that the events \eqref{eq:event-1}--\eqref{eq:event-4} are true---we can always choose a sufficiently small $\epsilon>0$ such that these events occur with arbitrarily high probabilities.

After source coding, the users now send $\{\mbf{c}_i, \mbf{c}_{ij}\}$ to the relay on the uplink. The common subcode known to all three users, i.e., $\mbf{c}_{123}$, need not be transmitted. Similar to FDF-IS, we will design channel codes for the relay to decode {\em functions} of the transmitted messages. %This scheme will exploit the fact that $\mbf{c}_{ij}$ is known to users $i$ and $j$.
This can be realized using linear block codes of the following form:
\begin{equation}
\mbf{x} = ( \mbf{c} \odot \mathbb{G}) \oplus \mbf{q}, \label{eq:linear-channel-code}
\end{equation}
where $\mbf{c} \in \mathcal{F}^\ell$ is the message vector, $\mathbb{G} \in \mathcal{F}^{\ell \times n}$ code generator matrix, $\mbf{q} \in \mathcal{F}^\ell$ is a random dither, and $\mbf{x} \in \mathcal{F}^n$ is the channel codeword. All elements are in $\mathcal{F}$, and $\odot$ is the multiplication in $\mathcal{F}$. Each element in $\mathbb{G}$ and in $\mbf{q}$ is independently and uniformly chosen over $\mathcal{F}$, and is known to the relay.

We now state the following lemma as a direct result of using linear block codes~\cite{ongmjohnsonit11}:
\begin{lemma} \label{lemma:linear-block-channel-code}
Each user $i$ transmits $\mbf{c}_i \in \mathcal{F}^\ell$ using the linear block code of the form \eqref{eq:linear-channel-code} with a common $\mathbb{G}$ and independently generated $\mbf{q}_i$ for each user. The relay receives $\mbf{Y}_0 = \mbf{X}_1 \oplus \mbf{X}_2 \oplus \mbf{X}_3 \oplus \mbf{N}_0 \in \mathcal{F}^n$ according to \eqref{eq:uplink}. If $n$ is sufficiently large and if
\begin{equation}
\frac{\ell \log F}{n} < \log F - H(N_0),
\end{equation}
then the relay can reliably\footnote{We say that a node can reliably decode a message if it can decode the message with arbitrarily small error probability.} decode the finite-field sum of the messages $\mbf{c}_1 \oplus \mbf{c}_2 \oplus \mbf{c}_3$.
\end{lemma}

From \eqref{eq:order}, \eqref{eq:subcode-length-1}--\eqref{eq:subcode-length-3}, and noting $H(W_i|W_j,W_k) + I(W_i;W_j|W_k) + H(W_j|W_i,W_k) =  H(W_i,W_j|W_k)$ we choose % choose a sufficiently large $m$ such that
\begin{equation}
\ell_1 + \ell_{12} + \ell_2 \geq \max \{ \ell_2 + \ell_{23} + \ell_3, \ell_1 + \ell_{13} + \ell_3\}. \label{eq:sub-code-length-compare}
\end{equation}

We consider the following two cases, where the relay decodes different functions in each case:

\subsubsection{When $\ell_{12} \leq \ell_3$\\ \indent\indent   (chosen when $I(W_1;W_2|W_3) \leq H(W_3|W_1,W_2)$) }$\;$

{\bfseries Uplink:}
We split the message $\mbf{c}_1$ into three different disjoint parts $\mbf{c}_1^{(1)}, \mbf{c}_1^{(2)},$ and $\mbf{c}_1^{(3)}$, and the message $\mbf{c}_2$ into $\mbf{c}_2^{(1)}, \mbf{c}_2^{(2)},$ and $\mbf{c}_2^{(3)}$. The uplink message transmission is arranged as shown in Table~\ref{table:1}.

The messages in each column are transmitted simultaneously using linear block codes with the message length specified in the first row and the codeword length in the second last row. From \eqref{eq:sub-code-length-compare}, we know that $\ell_1 + \ell_{12} - \ell_3 - \ell_{23} \geq 0$ and $\ell_2 + \ell_{12} - \ell_3 - \ell_{13} \geq 0$, meaning that the message length for each column is non-negative. For each column, both messages use the same code generator matrix but different dithers. The relay decodes the finite-field addition of the messages in each column. Take the first column for example, $\mbf{x}_1 = (\mbf{c}_1^{(1)} \odot \mathbb{G} ) \oplus \mbf{q}_a$ and $\mbf{x}_2 = (\mbf{c}_{23} \odot \mathbb{G}) \odot \mbf{q}_b$ are transmitted by user 1 and user 2 respectively. Note that the second codeword can also be transmitted by user 3 since it knows $\mbf{c}_{23}$. Using Lemma~\ref{lemma:linear-block-channel-code}, if $n$ is sufficiently large and if 
\begin{equation}
\frac{\ell_{23} \log F}{ \frac{\ell_{23}}{\ell_\text{T}}n} = \frac{\ell_\text{T} \log F}{n} < \log F - H(N_0), \label{eq:uplink-code-rate}
\end{equation}
where $\ell_\text{T} \triangleq \ell_1 + \ell_{12} + \ell_2$,
then the relay can reliably decode $\mbf{c}_1^{(1)} \oplus \mbf{c}_{23}$. Using the same coding scheme for the other columns, we can show that if \eqref{eq:uplink-code-rate} holds, then the relay can decode the summation of the messages in every column.

 {\bfseries Downlink:}
Assume that the relay has successfully decoded the functions for all columns $\mbf{c}_\text{T} = ( \mbf{c}_1^{(1)} \oplus \mbf{c}_{23},$ $\mbf{c}_1^{(3)},$ $\mbf{c}_1^{(2)} \oplus \mbf{c}_3^{(2)},$ $\mbf{c}_{12} \oplus \mbf{c}_3^{(1)},$ $\mbf{c}_2^{(2)} \oplus \mbf{c}_3^{(2)}$ $\mbf{c}_2^{(3)},$ $\mbf{c}_2^{(1)} \oplus \mbf{c}_{13})$. Note that $\mbf{c}_T$ is a finite-field vector of length $\ell_\text{T}$. Generate $F^{\ell_\text{T}}$ codewords of length $n$, where each codeletter is independently generated according to the uniform distribution $p^\text{u}(x_0)$. Index the codewords by $\mbf{x}_0(\mbf{c}_\text{T})$. After decoding $\mbf{c}_\text{T}$, the relay transmits $\mbf{x}_0(\mbf{c}_\text{T})$ on the downlink. By reducing the decoding space of each user---since each user has some side information about $\mbf{c}_\text{T}$---we can show the following (see, e.g., Ong and Johnson~\cite{ongjohnson12cl}):
\begin{lemma} \label{lemma:downlnk}
If user $i$ knows (a priori)  $\ell' \leq \ell_\text{T}$ elements in $\mbf{c}_\text{T}$, then it can reliably decode $\mbf{c}_\text{T}$ if $n$ is sufficiently large and if
\begin{equation}
\frac{(\ell_\text{T} - \ell') \log F}{n} < I(X_0;Y_i) \Big\vert_{p^\text{u}(x_0)} = \log F - H(N_i). 
\end{equation}
\end{lemma}
Note that random codes are used on the downlink instead of linear codes.

Knowing $\mbf{c}_1^{(3)}$ of length $\ell_1 + \ell_{12} - \ell_3 - \ell_{23}$, user 1 can reliably decode $\mbf{c}_\text{T}$ if
\begin{equation}
\frac{(\ell_2 + \ell_{23} + \ell_3) \log F}{n} < \log F - H(N_1). \label{eq:user-1-downlink}
\end{equation}
From $\mbf{c}_\text{T}$ and knowing its own subcodes $\mbf{c}_1^{(1)}, \mbf{c}_1^{(2)}, \mbf{c}_1^{(3)}, \mbf{c}_{12},$ and $\mbf{c}_{13}$, user 1 can then obtain  $\mbf{c}_2^{(1)},$ $\mbf{c}_2^{(2)},$ $\mbf{c}_2^{(3)},$ $\mbf{c}_{23},$ $\mbf{c}_3^{(1)},$ and $\mbf{c}_3^{(2)}$.

Knowing $\mbf{c}_2^{(3)}$ of length $\ell_2 + \ell_{12} - \ell_3 - \ell_{13}$, user 2 can reliably decode $\mbf{c}_\text{T}$ if
\begin{equation}
\frac{(\ell_1 + \ell_{13} + \ell_3) \log F}{n} < \log F - H(N_2). \label{eq:user-2-downlink}
\end{equation}
From $\mbf{c}_\text{T}$ and knowing its own subcodes $\mbf{c}_2^{(1)}, \mbf{c}_2^{(2)}, \mbf{c}_2^{(3)}, \mbf{c}_{12},$ and $\mbf{c}_{13}$, user 2 can then obtain $\mbf{c}_1^{(1)},$ $\mbf{c}_1^{(2)},$ $\mbf{c}_1^{(3)},$ $\mbf{c}_{13},$ $\mbf{c}_3^{(1)},$ and $\mbf{c}_3^{(2)}$.

Similarly, we can show that user 3 can reliably decode $\mbf{c}_\text{T}$ if
\begin{equation}
\frac{(\ell_1 + \ell_{12} + \ell_2) \log F}{n} < \log F - H(N_3). \label{eq:user-3-downlink}
\end{equation}
It can then proceed to obtain $\mbf{c}_1^{(1)}, \mbf{c}_1^{(2)}, \mbf{c}_1^{(3)}, \mbf{c}_{12},$ $\mbf{c}_2^{(1)}, \mbf{c}_2^{(2)}, \mbf{c}_2^{(3)}$.

{\bfseries Recovering Other Users' Messages:}
We have shown that from $\mbf{c}_\text{T}$, each user can obtain all other users' subcodes. If \eqref{eq:subcode-length-1}--\eqref{eq:subcode-length-4} are satisfied and if $m$ is sufficiently large, each user $i$ can reliably decode the messages of the other users, i.e., $\mbf{W}_j$ and $\mbf{W}_k$ from the subcodes.

{\bfseries Achievability:}
We now combine the above results. For any $\theta >0$, we can choose $\{\ell_i, \ell_{ij}\}$ (where $\ell_{12} \leq \ell_3$, which is possible because $I(W_1;W_2|W_3) \leq H(W_3|W_1,W_2)$) and a sufficiently large $m$ so that the lengths of the subcodes for source coding satisfy
\ifx\doublecolumn\undefined
\begin{align}
H(W_i|W_j,W_k) &< \frac{\ell_i \log F}{m} <  H(W_i|W_j,W_k) + \theta, && \textnormal{ for all } i\\
I(W_i;W_j|W_k) &< \frac{\ell_{ij} \log F}{m} < I(W_i;W_j|W_k) + \theta, && \textnormal{ for all } i < j.
\end{align}
\else
\begin{align}
H(W_i|W_j,W_k) &< \frac{\ell_i \log F}{m} <  H(W_i|W_j,W_k) + \theta, \nonumber \\ &\quad\quad\quad\quad\quad\quad\quad\quad\quad\quad\quad\quad \textnormal{ for all } i,\\
I(W_i;W_j|W_k) &< \frac{\ell_{ij} \log F}{m} < I(W_i;W_j|W_k) + \theta,  \nonumber \\ &\quad\quad\quad\quad\quad\quad\quad\quad\quad\quad \textnormal{ for all } i < j.
\end{align}
\fi
This means \eqref{eq:subcode-length-1}--\eqref{eq:subcode-length-3} are satisfied. It follows that
\ifx\doublecolumn\defined
\begin{align}
\frac{\kappa(\ell_i + \ell_{ij} + \ell_j)\log F}{n} &< H(W_i|W_j,W_k) + I(W_i;W_j|W_k) + H(W_j|W_i,W_k) + 3\theta, \\
\frac{(\ell_i + \ell_{ij} + \ell_j)\log F}{n} &<   \frac{H(W_i,W_j|W_k) + 3\theta}{\kappa}, \label{eq:small-theta}
\end{align}
\else
\begin{align}
\frac{\kappa(\ell_i + \ell_{ij} + \ell_j)\log F}{n} &< H(W_i|W_j,W_k) + I(W_i;W_j|W_k)\nonumber \\ &\quad  + H(W_j|W_i,W_k) + 3\theta, \\
\intertext{which simplifies to}
\frac{(\ell_i + \ell_{ij} + \ell_j)\log F}{n} &<   \frac{H(W_i,W_j|W_k) + 3\theta}{\kappa}, \label{eq:small-theta}
\end{align}
\fi
for all $i < j$.
The length $\ell_{123}$ is chosen to satisfy \eqref{eq:subcode-length-4}.

Now, for any $n/m \triangleq \kappa > \Phi$ in \eqref{eq:kappa-simplified}, i.e., $\kappa = \max \left\{ \frac{H(W_1,W_2|W_3)}{\log F - H(N_0)}, \max_{i,j,k} \frac{H(W_j,W_k|W_i)}{\log F - H(N_i)} \right\} + \theta'$ for some $\theta'>0$, we can choose a much smaller $\theta>0$ for \eqref{eq:small-theta}  such that \eqref{eq:uplink-code-rate}, \eqref{eq:user-1-downlink}, \eqref{eq:user-2-downlink}, and \eqref{eq:user-3-downlink} are simultaneously satisfied. 
With a sufficiently large $m$ (which also implies a large $n$), each user can reliably decodes the messages of the two other users. Hence, the source-channel rate $\kappa$ in \eqref{eq:kappa-simplified} is achievable. This proves the achievability of Theorem~\ref{theorem:main} for Case 3 when $\ell_{12} \leq \ell_3$.

\subsubsection{When $\ell_{12}  > \ell_3$ \\ \indent\indent   (chosen when $I(W_1;W_2|W_3) > H(W_3|W_1,W_2)$)}$\;$ \\
\indent The coding scheme for this case is similar to $\ell_{12} \geq \ell_3$. The uplink transmission is shown in Table~\ref{table:2}.

\begin{table}[h]
\centering
\caption{Uplink transmission using linear block codes when $\ell_{12}  > \ell_3$, $\ell_\text{T} \triangleq \ell_1 + \ell_{12} + \ell_2$}
\label{table:2}
\begin{tabular}{| c || c | c | c | c | c | c |}
\hline
Message length & $\ell_1$ & $\ell_3$ & \multicolumn{3}{c|}{$\ell_{12} - \ell_3$} & $\ell_2$ \\
\hline
Messages & $\mbf{c}_1$ & $\mbf{c}_{12}^{(1)}$ & \multicolumn{3}{c|}{ $\mbf{c}_{12}^{(2)}$} & $\mbf{c}_2$ \\
\hline
Messages &  & $\mbf{c}_3$ & \multicolumn{3}{c|}{} & \\
\hline
Messages & $\mbf{c}_{23}^{(1)}$ & & $\mbf{c}_{23}^{(2)}$ & \multicolumn{2}{c|}{ $\mbf{0}$} & \\
\hline
Messages & & & \multicolumn{2}{c|}{$\;\mbf{c}_{13}^{(2)}\;$} & $\mbf{0}$ & $\mbf{c}_{13}^{(1)}$ \\
\hline
Channel uses & $\frac{\ell_1}{\ell_\text{T}}n$ & $\frac{\ell_3}{\ell_\text{T}}n$ & \multicolumn{3}{c|}{ $\frac{\ell_{12}-\ell_3}{\ell_\text{T}}n$} & $\frac{\ell_2}{\ell_\text{T}}n$ \\
\hline
Total channel uses & \multicolumn{6}{c|}{$n$}\\
\hline
\end{tabular}
\end{table}

 The messages in each column are transmitted simultaneously. Note that in the third row of messages in the table, $\mbf{c}_{23}$ is split into two parts $(\mbf{c}_{23}^{(1)},\mbf{c}_{23}^{(2)})$ if and only if $\ell_{23} > \ell_1$. Else, the entire message $\mbf{c}_{23}$ will be transmitted in the first column, i.e., together with $\mbf{c}_1$. Since $\ell_1 + \ell_{12} + \ell_2 \geq \ell_3 + \ell_{23} + \ell_2 \Rightarrow \ell_{23} \leq \ell_1 + (\ell_{12} - \ell_3)$, the message $(\mbf{c}_{23}^{(1)},\mbf{c}_{23}^{(2)})$ can always fit into the first and third columns, with the remaining ``space'' padded with zero, denoted by $\mbf{0}$. The message $\mbf{c}_{13}$ is transmitted in a similar way.

The relay decodes the modulo addition of the messages in each column. If $n$ is sufficiently large and if \eqref{eq:uplink-code-rate} is satisfied, then the relay can reliably decode its intended messages, i.e., $\mbf{c}_\text{T} = ( \mbf{c}_1 \oplus \mbf{c}_{23}^{(1)}, \mbf{c}_{12}^{(1)} \oplus \mbf{c}_3, \mbf{c}_{12}^{(2)} \oplus (\mbf{c}_{23}^{(2)}, \mbf{0}) \oplus (\mbf{c}_{13}^{(2)}, \mbf{0}), \mbf{c}_{2} \oplus \mbf{c}_{13}^{(1)})$. The relay broadcasts $\mbf{c}_\text{T}$ on the downlink. Using Lemma~\ref{lemma:downlnk}, we can show that if  \eqref{eq:user-1-downlink}, \eqref{eq:user-2-downlink}, and \eqref{eq:user-3-downlink} are satisfied, then each user can reliably decode $\mbf{c}_\text{T}$, from which it can recover the messages of the other two users.

This completes the proof for the achievability of Case 3 in Theorem~\ref{theorem:main}. $\hfill \blacksquare$

\section{Discussions} \label{section:conclusion}

\subsection{Other Coding Schemes}

Note that the lower bound in Lemma~\ref{lemma:lower-bound} and the achievable source-channel rates in Lemma~\ref{lemma:achievability-separate} are  applicable to the general finite-field MWRC with correlated sources, in addition to Cases 1 and 2 in Theorem~\ref{theorem:main}. However, the coding technique in Section~\ref{section:upper-3} is useful only for sources that have their common information equal their mutual information.

Besides the coding schemes considered in this paper, one could treat the uplink (as a multiple-access channel with correlated sources) and the downlink (as a broadcast channel with receiver side information) separately to get potentially different achievable source-channel rates. In this case, on the uplink, we let the relay decode all three users' messages $(\mbf{W}_1, \mbf{W}_2, \mbf{W}_3)$. An achievable channel rate region for the two-sender multiple-access channel with correlated sources was found by Cover et al.~\cite{covergamal80}. Then, on the downlink, we can use the result by Tuncel~\cite{tuncel06} for the relay to transmit $(\mbf{W}_1,\mbf{W}_2,\mbf{W}_3)$ to the users taking into account that each user $i$ has side information $\mbf{W}_i$. % The combination of these two schemes gives a set of achievable rates. 

Extending the results of Cover et al.~\cite[Theorem 2]{covergamal80} to three senders, for the relay to be able to reliably decode $(\mbf{W}_1, \mbf{W}_2, \mbf{W}_3)$ on the uplink, $\kappa$ must satisfy the following:
\ifx\doublecolumn\undefined
\begin{align}
&& H(W_1, W_2, W_3) &< \kappa I(X_1,X_2,X_3;Y_0) \leq \kappa [\log F - H(N_0)], \\
\Rightarrow && \kappa &> \frac{H(W_1,W_2,W_3)}{\log F - H(N_0)}. \label{eq:cover-tuncel}
\end{align}
\else
\begin{align}
&& H(W_1, W_2, W_3) &< \kappa I(X_1,X_2,X_3;Y_0) \nonumber \\ && & \leq \kappa [\log F - H(N_0)], \\
\Rightarrow && \kappa &> \frac{H(W_1,W_2,W_3)}{\log F - H(N_0)}. \label{eq:cover-tuncel}
\end{align}
\fi
For the case where each user has non-zero message, $H(W_1,W_2,W_3) = H(W_i) + H(W_j,W_k|W_i) > H(W_j,W_k|W_i)$ for all $i$. Comparing \eqref{eq:cover-tuncel} to \eqref{eq:main-result}, we see that this coding strategy is strictly suboptimal for all Cases 1--3 in Theorem~\ref{theorem:main} when $H(N_0) \geq \max_{i \in \{1,2,3\}} H(N_i)$. However, this strategy {\em may} achieve better (i.e., lower) source-channel rates than SW/FDF-IS in general. We leave the derivation of the rates to the reader---it is straightforward given the results by Cover et al.\ and Tuncel.

In this paper, we have only investigated the separate-source-channel coding architecture without feedback. Though we have identified source/channel combinations where the minimum source-channel rate is found, the problem remains open in general. Two directions which one could explore are {\em joint} source-channel coding and {\em feedback}.

\subsection{Optimality of Source-Channel Separation}

We now give a numerical example where none of the coding schemes in this paper achieves the lower bound $\Phi$. Let the sources be $W_1 = V_1 \boxplus  V_{12} \boxplus V_{13}$, $W_2 = V_2 \boxplus V_{12} \boxplus V_{23}$, and $W_3 = V_3 \boxplus V_{13} \boxplus V_{23}$ where all $V_{\dotsc} \in \{0,1\}$ are independent random variables, and $\boxplus$ denotes modulo-two addition. We choose $\Pr\{V_1 = 1\} = 0.3$, $\Pr\{V_2 = 1\} = \Pr\{V_3 =1 \} = 0.029$, $\Pr\{ V_{12} = 1\} = \Pr\{V_{13} = 1\} = 0.103$, and $\Pr\{V_{23} = 1\} = 0.33$. For this choice, we have $H(W_1|W_2,W_3) = 0.90$, $H(W_2|W_1,W_3) = H(W_3|W_1,W_2) = 0.70$, $H(W_1,W_2|W_3) = H(W_1,W_3|W_2) = H(W_2,W_3|W_1) = 1.65$.

For the channel, let $\mathcal{F} = \{0,1,2,3\}$, which gives $\log F = 2$. We choose $\Pr\{N_0 = 0 \} = 1$ and $\Pr\{N_0 = a\} = 0 $ for $a \in \{1,2,3\}$, giving $H(N_0) = 0$. For each $i \in \{1,2,3\}$, we set $\Pr\{N_i = 0\} = 0.0533$, $\Pr\{N_i=1\} = 0.9467$, and $\Pr\{N_i =a \}= 0$ for $a \in \{2,3\}$, giving $H(N_i) = 0.30$. This means, $\log F - H(N_0) = 2$, and $\log F - H(N_i) = 1.70$ for all $i \in \{1,2,3\}$.

In this example, $\Phi = 0.9705$. If $\kappa = 0.9$ is achievable, then (from Lemma~\ref{lemma:achievability-separate}) we must be able to find some $R_2$ and $R_3$ satisfying $R_2 + R_3 > H(W_2,W_3|W_1)/\kappa = 1.833$, and $R_2 + R_3 < \log F - \max\{ H(N_0), H(N_1)\} = 1.7$. Since the conditions cannot be simultaneously met, $\kappa = 0.9$ is not achievable using SW/FDF-IS. As the sources' common information does not equal their mutual information, we cannot use the coding scheme derived in Section~\ref{section:upper-3}. 

This example shows that the separation schemes derived in this paper cannot achieve the lower bound in some cases. However, to show that separation is suboptimal, one has to explore all possible separation schemes and show that some joint source-channel scheme achieves a better source-channel rate.

\subsection{Examples of Sources in Cases 1 and 3}

In this paper, we have identified three classes of source/channel combinations where the minimum source-channel rate is found. The first class is sources that have almost-balanced conditional mutual information (ABCMI). An example of sources that have ABCMI is interchangeable random variables in the sense of Chernoff and Teicher~\cite{chernoffteicher58}, where ``every subcollection of the random variables has a joint distribution which is a symmetric function of its arguments.'' This can model sensor networks where the sensors are equally-spaced on a circle to detect a phenomenon occurring at the center of the circle.
 However, the ABCMI conditions are looser than that of interchangeable random variables as the former only requires that mutual information between any two sources has {\em roughly} the same value (see Appendix~\ref{appendix:abcmi}), and also, the marginal distribution of the variables can be vastly different.

Another class for which we have derived the minimum source-channel rate is when the sources have their common information equal their mutual information. An example is correlated sources in the sense of Han~\cite{han79}, where the sources can be written as $W_1 = ( U_1, U_{12}, U_{13}, U_{123} )$, $W_2 = ( U_2, U_{12}, U_{23}, U_{123} )$, and $W_3 = ( U_3, U_{13}, U_{23}, U_{123} )$, where $U_1, U_2, U_3,$ $U_{12}, U_{13}, U_{23},$ and $U_{123}$ are {\em mutually independent} random variables. %, i.e., $\Pr\{P=p, Q=q, R=r, S=s, T=t, U=u, V=v\}= \Pr\{P=p\}\Pr\{Q=q\}\Pr\{R=r\}\Pr\{S=s\}\Pr\{T=t\}\Pr\{U=u\}\Pr\{V=v\}$. 
Using sensor networks as an example again, each node here has multiple sensing capabilities, e.g., temperature, light, sound. As these measurements display different behavior spatially, some remain constant across subsets of sensors, e.g., nodes 1 and 2 always sense the same temperature but different light intensity.

%As for the class of sources with skewed conditional entropies, the conditions are purely mathematical.

As for the class of sources with skewed conditional entropies, the conditions appear to be purely mathematical in nature.

\section*{Acknowledgment}
\addcontentsline{toc}{section}{Acknowledgment}

The authors would like to thank Roy Timo for discussions on G{\'a}cs and K{\"o}rner's common information and other helpful comments.

\appendices

\section{Graphical Interpretation of Sources that Have ABCMI and SCE} \label{appendix:abcmi}

\begin{figure}[t]
\centering
\subfigure[almost-balanced conditional mutual information (ABCMI)]{
\includegraphics[width=6.5cm]{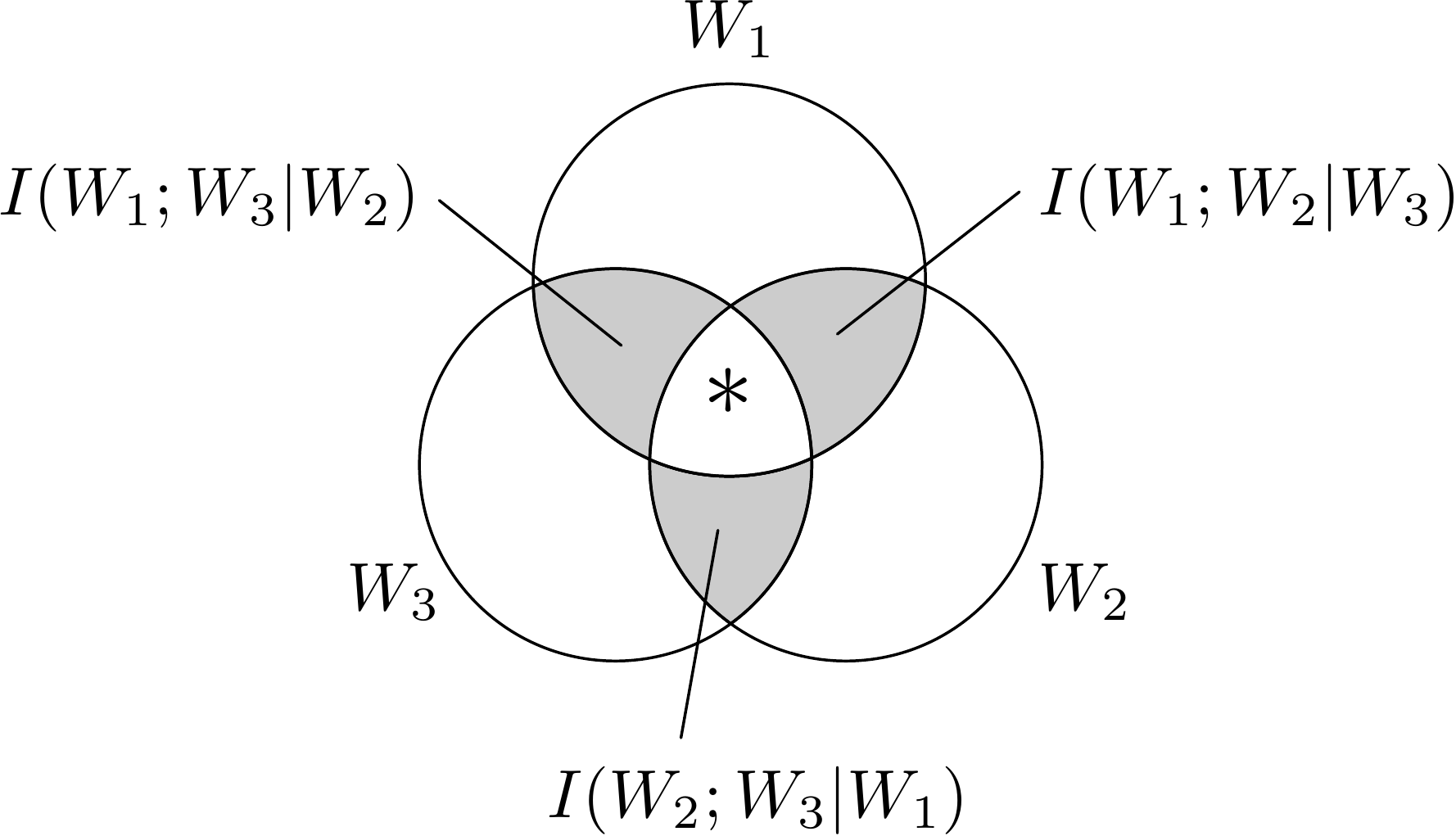}
\label{fig:case-1}
}
\hspace{1cm}
\subfigure[skewed conditional entropies (SCE)]{
\includegraphics[width=7.5cm]{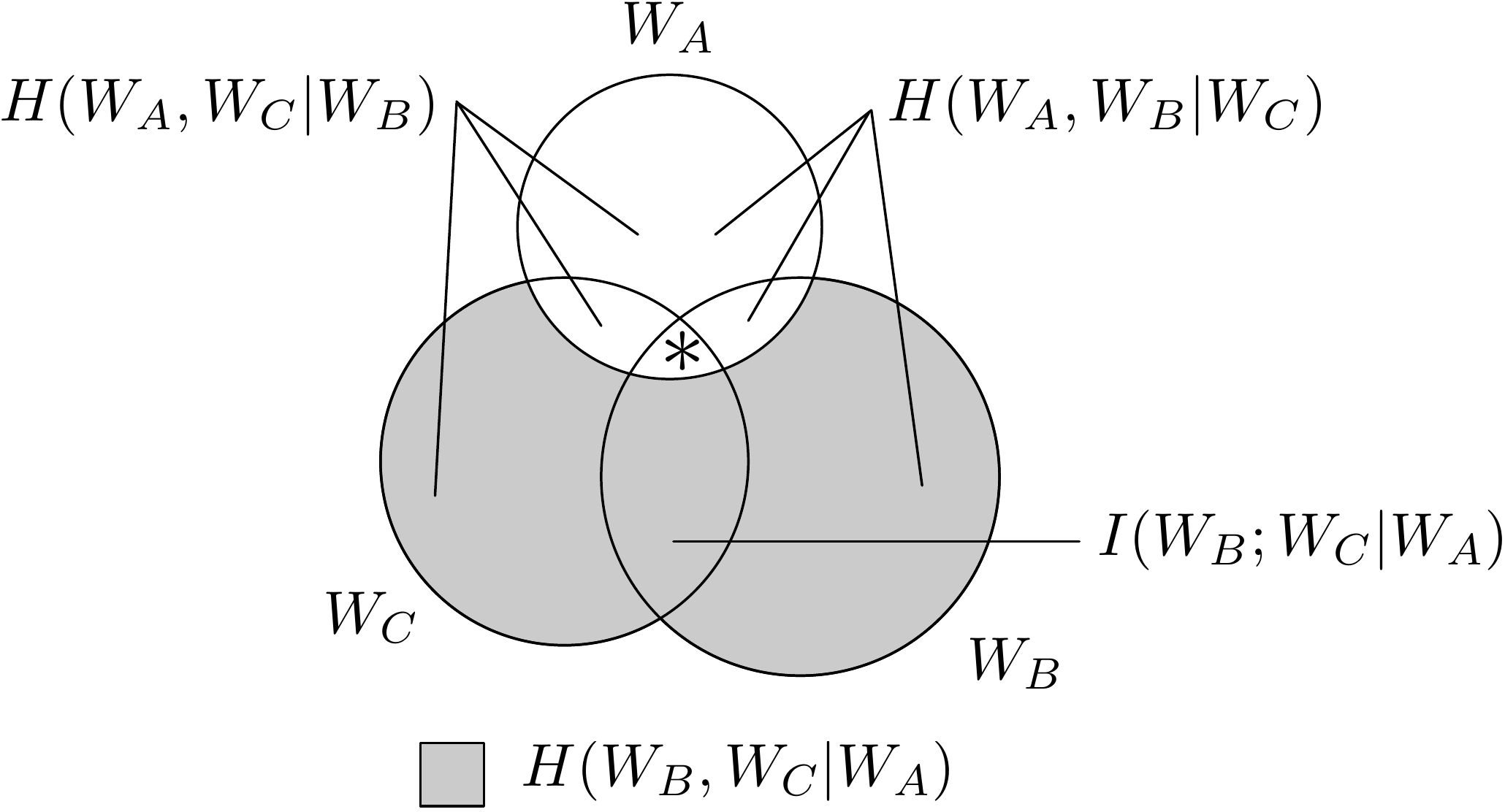}
\label{fig:case-2a}
}
% \subfigure[unbalanced conditional mutual information and without SCE]{
% \includegraphics[width=3.5cm]{../pictures/mwrc-cs-pic-08}
% \label{fig:case-2b}
% }
\caption{Entropy diagrams of three correlated sources: the values of $I(\cdot)$ and $H(\cdot)$ can be represented as signed measures on each diagram; the measures are drawn proportional to the areas (except for the measure of the region $*$, which can be negative)}
\label{fig:case1-2}
\end{figure}

Figure~\ref{fig:case1-2} shows the relationship among the entropies and mutual information for the three source messages $W_1$, $W_2$, and $W_3$ for the cases described above. Referring to Figure~\ref{fig:case-1}, the shaded areas represent the mutual information between any two source messages given the third source message. For ABCMI, we have that any of the three shaded areas must not be bigger than the sum of the other two shaded areas. Suppose that the sources do not have ABCMI, then they must have unbalanced conditional mutual information, i.e., we can find a user $A$ where $I(W_B;W_C|W_A)$ is larger than the sum of $I(W_A;W_B|W_C)$ and $I(W_A;W_C|W_B)$ by an amount $\eta>0$ (see Figure~\ref{fig:case-2a}). In addition, for sources that have SCE, we also have that for the two messages, $W_B$ and $W_C$, whose mutual information conditioned on $W_A$, i.e., $I(W_B;W_C|W_A)$, is larger than the sum of the other two pairs by the amount $\eta$, their entropy conditioned on $W_A$, i.e., $H(W_B,W_C|W_A)$, is also greater than that of any other pair (conditioned on the message of the third user) by at least $\eta$. The information diagram for SCE is depicted in Figure~\ref{fig:case-2a}.

\section{Proof of Lemma~\ref{lemma:achievability-separate}} \label{appendix:slepian-wolf-fdf}

We first quote two existing results of (i) channel coding for the three-user MWRC with independent messages and (ii) source coding with side information. %These results will be used in proving cases 1 and 2 in Theorem~\ref{theorem:main}. %We use upper-case letters $R_i$ to denote channel coding rates, and lower-case letters $r_i$ to denote source coding rates.

\subsection{Functional-Decode-Forward for Independent Sources (FDF-IS) Channel Coding} \label{sec:channel-coding}

The following channel-coding setting assumes that the source messages are independent:
\begin{lemma}[Ong et al.~\cite{ongmjohnsonit11}] \label{lemma:channel-coding}
Consider the MWRC depicted in Figure~\ref{fig:mwrc}, where each user $i$'s message $W_i$ is uniformly distributed in $\{1,2,\dotsc, M_i\}$ (we consider a single copy, i.e., $m=1$), and where $W_1$, $W_2$, and $W_3$ are independent. 
The channel is used $n$ times according to the block code (with $m=1$) specified in Section~\ref{sec:model}. Each user can then decode the messages of the other two users (with its $n$ received channel symbols and its own message) with an arbitrarily small error probability if $n$ is sufficiently large, and if  %, i.e., reliable communication is possible,
\begin{equation}
\frac{\log M_i + \log M_j}{n} < \log F - \max \{ H(N_0), H(N_i) \} , \label{eq:channel-coding}
\end{equation}
for all distinct $i,j,k \in \{1,2,3\}$.
\end{lemma}

\subsection{Slepian-Wolf Source Coding} \label{sec:source-coding}
The following source-coding setting assumes that the channel is noiseless:

\begin{lemma}[Cover~\cite{cover75}] \label{lemma:source-coding}
Consider only the three users with their respective length-$m$ messages $\{\mbf{W}_i\}$ generated according to (\ref{eq:source}). Each user $i$ encodes its messages $\mbf{W}_i$ to an index $V_i \in \{1,2,\dotsc, M_i\}$, and gives its index to the other two users. Each user can then decode the messages of the other two users (with the received indices and its own message) with an arbitrarily small error probability if $m$ is sufficiently large, and if
\begin{align}
\frac{\log M_i}{m} &>   H(W_i|W_j,W_k) , \label{eq:source-coding-1} \\
\frac{\log M_i + \log M_j}{m} &>  H(W_i,W_j|W_k), \label{eq:source-coding-2}
\end{align}
for all distinct $i,j,k \in \{1,2,3\}$.
\end{lemma}

Note that Wyner et al.~\cite{wynerwolf02} derived a similar result with an additional constraint on the relay. In their setup, the users present their indices to a relay; the relay in turn re-encodes and presents its index to the users.

\subsection{Proof of Lemma~\ref{lemma:achievability-separate}}
We use Slepian-Wolf source coding. Each user $i$ encodes its length-$m$ message $\mbf{W}_i$ to an index $V_i' \in \{0,1,\dotsc, M_i-1\}$, satisfying \eqref{eq:source-coding-1} and \eqref{eq:source-coding-2}, for $i \in \{1,2,3\}$. Each user $i$ randomly generates a dither $D_i$ uniformly distributed in $\{0,1,\dotsc, M_i-1\}$, and forms its encoded message $V_i = V_i' + D_i \mod M_i$. The dithers are made known to all nodes. Now, $V_1$, $V_2$, and $V_3$ are mutually independent, and each $V_i$ is uniformly distributed in $\{0,1,\dotsc, M_i-1\}$.

We then use FDF-IS channel coding for the users to exchange the encoded independent messages $V_1$, $V_2$, and $V_3$ via the relay in $n$ channel uses. From Lemma~\ref{lemma:channel-coding}, if \eqref{eq:channel-coding} is satisfied, then each user $i$ can reliably recover $V_j$ and $V_k$. Knowing the dithers, it can also recover $V_j'$ and $V_k'$.
From Lemma~\ref{lemma:source-coding}, if \eqref{eq:source-coding-1} and \eqref{eq:source-coding-2} are satisfied, then each user $i$ can reliably recover $(\mbf{W}_j,\mbf{W}_k)$.

Noting that $\kappa \triangleq n/m$ and defining $R_i = \log M_i/n$, the conditions for achievability, i.e., \eqref{eq:channel-coding}, \eqref{eq:source-coding-1}, and \eqref{eq:source-coding-2}, can be expressed as \eqref{eq:achievability-separate}. $\hfill \blacksquare$

\section{Coding for Sources That Have Their Common Information Equal Their Mutual Information} \label{app:GK-source-coding}

We perform source coding for correlated sources~\cite{slepianwolf73}\cite{cover75}. Consider the source message $\boldsymbol{W}_1$. Clearly, $H(V_{12}|W_1) = H(V_{13}|W_1) = H(V_{123}|W_1) = 0$, since $V_{\{1..\}}$ are deterministic functions of $W_1$. Now since $V_{12}$ captures all information that $W_1$ and $W_2$ have in common (because $H(V_{12})=I(W_1;W_2)$), we have $H(V_{123}|V_{12}) = 0$. Similarly $H(V_{123}|V_{13}) = 0$.

So, we can reliably reconstruct $\boldsymbol{W}_1$ from $(\boldsymbol{c}_1, \boldsymbol{c}_{12}, \boldsymbol{c}_{13}, \boldsymbol{c}_{123})$ if $m$ is sufficiently large, and if the following inequalities hold~\cite[Theorem\ 2]{cover75}:
\ifx\doublecolumn\undefined
\begin{align*}
(\ell_{1} \log F )/m &> H(W_1|V_{12}, V_{13}, V_{123}) = H(W_1|W_2,W_3), \\
(\ell_{12} \log F) /m &> H(V_{12}|W_1, V_{13}, V_{123}) = 0, \\
(\ell_{13} \log F) /m &> H(V_{13}|W_1,V_{12}, V_{123}) = 0, \\
(\ell_{123} \log F)/m &> H(V_{123}|W_1, V_{12}, V_{13}) = 0,\\
([\ell_{1} + \ell_{12}] \log F ) /m &> H(W_1, V_{12} | V_{13}, V_{123}) = H(W_1|W_2,W_3) + I(W_1;W_2|W_3), \\
([\ell_{1} + \ell_{13}] \log F ) /m &> H(W_1, V_{13} |V_{12}, V_{123}) = H(W_1|W_2,W_3) + I(W_1;W_3|W_2), \\
([\ell_{1} + \ell_{123}] \log F) /m &> H(W_1, V_{123} | V_{12}, V_{13}) = H(W_1|W_2,W_3), \\
([\ell_{12} + \ell_{13}] \log F) /m &> H(V_{12}, V_{13} | W_1, V_{123}) = 0, \\
&\vdots\\
([\ell_{1} + \ell_{12} + \ell_{13}] \log F ) /m &> H(W_1, V_{12}, V_{13} | V_{123}) \nonumber \\ &= H(W_1|W_2,W_3) + I(W_1;W_2|W_3) + I(W_1;W_3|W_2), \\
([\ell_{1} + \ell_{12} + \ell_{123}] \log F) /m &> H(W_1, V_{12}, V_{123} | V_{13}) = H(W_1|W_2,W_3) + I(W_1;W_2|W_3), \\
([\ell_{1} + \ell_{13} + \ell_{123}] \log F) /m &> H(W_1, V_{13}, V_{123} | V_{12}) = H(W_1|W_2,W_3) + I(W_1;W_3|W_2), \\
([\ell_{12} + \ell_{13} + \ell_{123}\ \log F ) /m &> H(V_{12}, V_{13}, V_{123}|W_1) = 0, \\
([\ell_{1} + \ell_{12} + \ell_{13} + \ell_{123}] \log F) /m & = H(W_1, V_{12}, V_{13}, V_{123}) = H(W_1).
\end{align*}
\else
\begin{align*}
(\ell_{1} \log F )/m &> H(W_1|V_{12}, V_{13}, V_{123}) \\ &= H(W_1|W_2,W_3), \\
(\ell_{12} \log F) /m &> H(V_{12}|W_1, V_{13}, V_{123}) = 0, \\
(\ell_{13} \log F) /m &> H(V_{13}|W_1,V_{12}, V_{123}) = 0, \\
(\ell_{123} \log F)/m &> H(V_{123}|W_1, V_{12}, V_{13}) = 0,\\
([\ell_{1} + \ell_{12}] \log F ) /m &> H(W_1, V_{12} | V_{13}, V_{123})\\ & = H(W_1|W_2,W_3)\\ &\quad + I(W_1;W_2|W_3), \\
([\ell_{1} + \ell_{13}] \log F ) /m &> H(W_1, V_{13} |V_{12}, V_{123})\\ & = H(W_1|W_2,W_3)\\ &\quad  + I(W_1;W_3|W_2), \\
([\ell_{1} + \ell_{123}] \log F) /m &> H(W_1, V_{123} | V_{12}, V_{13})\\ & = H(W_1|W_2,W_3), \\
([\ell_{12} + \ell_{13}] \log F) /m &> H(V_{12}, V_{13} | W_1, V_{123}) = 0, \\
&\vdots\\
([\ell_{1} + \ell_{12} + \ell_{13}] \log F ) /m &> H(W_1, V_{12}, V_{13} | V_{123}) \nonumber \\ &= H(W_1|W_2,W_3)\\ &\quad  + I(W_1;W_2|W_3)\\ &\quad  + I(W_1;W_3|W_2), \\
([\ell_{1} + \ell_{12} + \ell_{123}] \log F) /m &> H(W_1, V_{12}, V_{123} | V_{13})\\ & = H(W_1|W_2,W_3)\\ &\quad  + I(W_1;W_2|W_3), \\
([\ell_{1} + \ell_{13} + \ell_{123}] \log F) /m &> H(W_1, V_{13}, V_{123} | V_{12})\\ & = H(W_1|W_2,W_3)\\ &\quad  + I(W_1;W_3|W_2), \\
([\ell_{12} + \ell_{13} + \ell_{123}\ \log F ) /m &> H(V_{12}, V_{13}, V_{123}|W_1) = 0,
\end{align*}
\begin{align*}
([\ell_{1} + \ell_{12} + \ell_{13} + \ell_{123}] \log F) /m & = H(W_1, V_{12}, V_{13}, V_{123})\\ & = H(W_1).
\end{align*}
\fi
Here, we need to consider all possible non-empty subsets of $\{\ell_{1}, \ell_{12}, \ell_{13}, \ell_{123}\}$ on the LHS. We have omitted some trivial inequalities where the RHS equals zero. Repeating the above for sources 2 and 3, and simplifying, we have \eqref{eq:subcode-length-1}--\eqref{eq:subcode-length-4}.

%\bibliography{../bib}


\end{document}